\documentclass[a4paper,fleqn,usenatbib]{mnras}

\usepackage{newtxtext,newtxmath}
\usepackage[T1]{fontenc}
\usepackage{ae,aecompl}

\usepackage{graphicx}	% Including figure files
\usepackage{amsmath}	% Advanced maths commands
\usepackage{amssymb}	% Extra maths symbols

\def\asec{\ifmmode ^{\prime\prime}\else$^{\prime\prime}$\fi}
\def\etal{{et\,al. }}

\def\msun{\hbox{M$_{\odot}$}}

\def\Msun{\hbox{~{\rm M}_\odot}}

\def\degs{\ifmmode ^{\circ}\else$^{\circ}$\fi}
\def\amin{\ifmmode ^{\prime}\else$^{\prime}$\fi}
\def\asec{\ifmmode ^{\prime\prime}\else$^{\prime\prime}$\fi}
\def\fdg{\hbox{$.\!\!^\circ$}}          % Fractions of degrees
\def\farcs{\hbox{$.\!\!^{\prime\prime}$}}  % Fractions of arcseconds
\def\degs{\ifmmode ^{\circ}\else$^{\circ}$\fi}
\def\amin{\ifmmode ^{\prime}\else$^{\prime}$\fi}

\def\EE#1{\times 10^{#1}}

\def\cm{\mbox{\,cm}}

\def\cm3{\mbox{\,cm$^{-3}$}}
\def\kms{\mbox{\,km~s$^{-1}$}}

\def\kms{\mbox{\,km s$^{-1}$}}

\def\lsim{\!\!\!\phantom{\le}\smash{\buildrel{}\over
 {\lower2.5dd\hbox{$\buildrel{\lower2dd\hbox{$\displaystyle<$}}\over
                                 \sim$}}}\,\,}
\def\gsim{\!\!\!\phantom{\ge}\smash{\buildrel{}\over
{\lower2.5dd\hbox{$\buildrel{\lower2dd\hbox{$\displaystyle>$}}\over
                               \sim$}}}\,\,}
\def\Msun{~{\rm M}_\odot}

\def\psr{PSR0540}
\def\pwn{PWN0540}
\def\snr{SNR~0540-69.3}                               
\title[Atacama Compact Array Observations of PWN 0540-69.3]{Atacama Compact Array Observations of the Pulsar-Wind Nebula of SNR 0540-69.3}

\author[P. Lundqvist et al.]{
P. Lundqvist$^{1,2}$\thanks{E-mail: peter@astro.su.se (PL)},
N. Lundqvist$^{1}$, C. Vlahakis$^{3}$, C.-I. Bj\"ornsson$^{1}$, J. R. Dickel$^{4}$, 
\newauthor M. Matsuura$^{5}$,  Yu. A. Shibanov$^{6,7}$, D. A. Zyuzin$^{6}$, G. Olofsson$^{1}$
\\
$^{1}$Department of Astronomy, AlbaNova University Center, Stockholm University, SE-10691 Stockholm, Sweden\\
$^{2}$The Oskar Klein Centre, AlbaNova, SE-10691 Stockholm, Sweden\\
$^{3}$National Radio Astronomy Observatory, 520 Edgemont Road, Charlottesville, VA 22903-2475, USA\\
$^{4}$Department of Astronomy, University of Illinois Urbana-Champaign, 1002 W. Green Street, Urbana, IL 61801, USA\\
$^{5}$School of Physics and Astrophysics, Cardiff University, Queens Buildings, The Parade, Cardiff CF24 3AA, UK\\
$^{6}$Ioffe Institute, Politekhnicheskaya 26, St. Petersburg, 194021, Russia\\
$^{7}$Peter the Great St. Petersburg Polytechnic University, Politekhnicheskaya 29, St. Petersburg, 195251, Russia
}

\date{Accepted XXX. Received YYY; in original form ZZZ}

\pubyear{2019}

\begin{document}
\label{firstpage}
\pagerange{\pageref{firstpage}--\pageref{lastpage}}
\maketitle

% Abstract of the paper
\begin{abstract}
% not more than 250 words (200 words for Letters).
%No references should appear in the abstract.
We present observations of the pulsar-wind
nebula (PWN) region of  \snr. The observations were made with the Atacama 
Compact Array (ACA) in Bands 4 and 6. We also add radio
observations from the Australia Compact Array (ATCA) at 3 cm. 
For $1.449 - 233.50$~GHz we 
obtain a synchrotron spectrum $F_{\nu} \propto \nu^{-\alpha_{\nu}}$, with
the spectral index $\alpha_{\nu} = 0.17\pm{0.02}$. To conclude
how this joins the synchrotron spectrum at higher frequencies we include hitherto
unpublished AKARI mid-infrared data, and evaluate published 
data in the ultraviolet (UV), optical and infrared (IR). In particular,
some broad-band filter data in the optical must be discarded from our
analysis due to contamination by spectral line emission. For the UV/IR part 
of the synchrotron spectrum, we arrive at $\alpha_{\nu} = 0.87^{+0.08}_{-0.10}$.
There is room for $2.5\times10^{-3} \Msun$ of dust with temperature $\sim 55$~K if 
there are dual breaks in the synchrotron spectrum, one around 
$\sim 9\EE{10}$~Hz, and another at $\sim 2\EE{13}$~Hz. The spectral index
then changes at $\sim 9\EE{10}$~Hz from $\alpha_{\nu} = 0.14\pm0.07$ in the radio, 
to $\alpha_{\nu} = 0.35^{-0.07}_{+0.05}$ in the millimetre to far-IR range. The ACA 
Band 6 data marginally resolves the PWN. In particular, 
the strong emission $\sim 1\farcs5$ south-west of the 
pulsar, seen at other wavelengths, and resolved in the 3-cm data with 
its 0\farcs8 spatial resolution, is also strong in the millimeter range.
The ACA data clearly reveal the supernova
remnant shell $\sim 20-35 \arcsec$ west of the pulsar, and for the shell we
derive $\alpha_{\nu} = 0.64\pm{0.05}$ for the range $8.6-145$~GHz.
\end{abstract}

% Select between one and six entries from the list of approved keywords.
% Don't make up new ones.
\begin{keywords}
pulsars: individual: PSR B0540-69.3 -- ISM: supernova remnants -- ISM: individual: SNR 0540-69.3 -- supernovae: general -- Magellanic Clouds
\end{keywords}

%%%%%%%%%%%%%%%%%%%%%%%%%%%%%%%%%%%%%%%%%%%%%%%%%%

%%%%%%%%%%%%%%%%% BODY OF PAPER %%%%%%%%%%%%%%%%%%

\section{Introduction}
The oxygen-rich ejecta of the $\sim 10^3$ year old LMC supernova remnant
(SNR) 0540-69.3 (in short SNR0540) \citep{Kirshner89,ser05} is evidence
that SNR0540 is the result of an explosion of a massive, $\sim20 \Msun$ 
(at zero-age main-sequence),  progenitor
\citep{Chevalier06,Williams08,nlun11}. 
The radiation emitted from the remnant manifests itself mainly through
radio and X-ray emission from a shell of radius $\sim20\arcsec-35\arcsec$,
corresponding to $\sim 4.9-8.5$~pc at the 50 kpc distance of LMC
\citep{man93b,GW00}. Most of this emission comes from the western part of
the shell.

The remnant contains the pulsar PSR B0540-69.3 (henceforth \psr),
discovered as a pulsed ($P = 50.2$~ms) X-ray source by \citet{sew84}.
Pulsations have subsequently also been detected in the optical
\citep{Middleditch85}, at radio wavelengths \citep{man93a}, and recently 
in the UV \citep{Mignani19}. The properties of \psr~are very similar to
those of the Crab pulsar: it spins rapidly and it is young (spin down age
1660 yr). 

Like the Crab pulsar, \psr~powers a pulsar-wind nebula (PWN), which we 
will refer to as \pwn. The full size of \pwn\ in radio is 
$\sim 7\arcsec-8\arcsec$ across, with an inner core of $\sim 5\arcsec$ in
diameter, also seen in the UV/optical \citep{Car92,serf04} radio
\citep[e.g.,][]{man93b} and X-rays \citep{GW00}.
The PWN emits synchrotron emission with a flux 
$F_{\nu} \propto \nu^{- \alpha_{\nu}}$, where $\alpha_{\nu}$ runs from
$\sim 0.15$ in the radio \citep{bra14} to a larger value in the optical
\citep{serf04,Mignani12}. 

\pwn\ is in no sense different from other 
PWNe when it comes to the shallow slope in radio; the vast majority of
39 PWNe studied in the radio have $\alpha_{\nu} \lsim 0.3$ \citep{rey17}.  
Neither is the steepening towards higher photon energies unique. For
example, the Crab PWN has a spectral break around $\sim 3\EE{10}$~Hz, 
where the spectral index changes from $\alpha_{\nu} \approx 0.3$ in the
radio to $\alpha_{\nu} \approx 0.42$ at higher frequencies \citep{gom12}.
Such spectral breaks have also been seen in other PWNe 
\citep[cf.][and references therein]{rey17}.

The shape of the synchrotron radio/IR spectrum provides important
information about the energy distribution of relativistic electrons and their 
radiative losses, and inhomogeneities 
within the PWN \citep{rey17}.  For a power-law distribution for the energy of 
relativistic electrons, $dN/dE = N_0E^{-p}$, where $E=\gamma m_ec^2$ is the 
energy of the electrons and $\gamma$ is the Lorentz factor, the intensity of optically 
thin synchrotron emission is $\propto \nu^{-\alpha}$, where $\alpha= (p-1)/2$. 
A flat spectrum with $\alpha \sim 0.15$, as for PWN0540 in the radio, would then 
indicate $p \sim 1.3$. This is much shallower than $p \gsim 2$ expected at shock 
fronts of relativistic shocks \citep[e.g.,][]{byk12}, corresponding to $\alpha \gsim 0.5$.
\citet{bucc11} showed that a broken power-law energy distribution of the
relativistic electrons can explain the shallow radio spectrum of the Crab PWN and 
a few other PWNe, and argued for that the relativistic electrons at the lowest energies 
could be accelerated in a turbulence zone in connection with the termination
shock rather than via Fermi acceleration. In the work of \citet{bucc11}, as well
of others \citep[e.g.,][]{Gelf09,Mart12}, particles are continuously injected 
into the PWN, with the energy spectrum of the injected particles being discontinuous or 
having a break.

There are models invoking reconnection as the source of particle 
acceleration \citep[e.g.,][]{Clau12,Cer12,Sir11,Porth16},
and \citet{Lyu19} argue that both Fermi-I and reconnection processes occur in the 
Crab PWN, and that this can explain the multiple spectral breaks for that PWN. 
Several spectral breaks are also evident in compiled IR-optical-X-ray 
spectra of the torus-like parts of other well-studied PWNe \citep[\pwn, 3C~58, J1124-5916 
and G21.5-0.9,][]{Zha13}, implying that multiple populations of relativistic particles are responsible 
for the emission in different spectral domains. In particular, 
the overall radio-to-X-ray spectrum of \pwn\ has been discussed by 
\citet{serf04}, \citet{Mignani12}, \citet{nlun11} and \citet{bra14}, and it shows that the 
X-ray emission is stronger than expected from a power-law extrapolation of the 
IR/optical/UV spectrum. This indeed indicates the existence of 
more than one emission component. In addition, spatially resolved X-ray spectra of \pwn\
show that the steepness of the X-ray spectrum increases away from the torus region \citep{nlun11}, 
which is expected if the particles injected there experience adiabatic and synchrotron losses 
as they move away from that region. Such spatial variations tend to smooth
spectral breaks in spatially integrated spectra of PWNe \citep[e.g.,][]{rey17}.

An established synchrotron contribution in the millimetre/IR range is needed to 
estimate possible additional continuum emission from supernova-produced
dust. Prime examples of the latter in SNRs are the Crab Nebula 
\citep{gom12,owen15}, PWN G54.1+0.3 \citep{tem17,rho18}, Cas A
\citep{delooze17}, and the young ($\lsim 2.5$ kyr) 
remnants G11.2-0.3, G21.5-0.9 and G29.7-0.3 \citep{chaw19}. The most
obvious  case outside the Milky Way is SN~1987A
\citep{mat11,mat15,ind14,dwek15}, which, like the SNRs mentioned, has 
substantial amounts of dust ($\gsim 0.2 \Msun$) embedded within it. 

Spitzer observations indicate that dust exists also in \pwn. The estimated
temperature and mass of this dust is $\sim 50-65$~K and 
$(1-3)\EE{-3} \Msun$, respectively \citep{Williams08}. These estimates 
are, however, sensitive to the shape of the underlying synchrotron 
spectrum in the IR. \citet{Williams08} used the results of 
\citet{serf04} in the UV and optical to extrapolate to the IR. A critical
assessment of the UV/optical synchrotron component is therefore essential 
to derive a reliable dust mass. The more recent estimate of by 
\citet{Mignani12} for the UV/optical/near-IR synchrotron 
spectrum indicates a notably higher flux than that of \citet{serf04}, and 
the derived synchrotron slope is markedly shallower. 
If the results of \citet{Mignani12} were to be combined with the Spitzer 
data, this would seriously affect the derived mass and
temperature of the dust discussed by \citet{Williams08}.

To better constrain the continuum emission from \pwn, we have obtained 
data from the Atacama Compact Array (ACA) in Bands 4 and 6, i.e., in the frequency range 
$137.0-233.5$ GHz\footnote{ALMA Program 2017.1.01391.S, PI: P. Lundqvist} 
(cf. Tables~\ref{tab:observations} and ~\ref{tab:observations2}). 
We have also complemented the millimetre ACA data with the high-resolution radio data of
\citet{dick02} to model the partially resolved ACA data. 
The reason for this modelling is the uneven spatial distribution of 
emission seen in the PWN at many wavelengths. In particular, 
much of the emission, especially in X-rays, but also in the optical, comes 
from a region $\sim 1\farcs5$ southwest of the pulsar, 
where a bright time-variable structure (``blob'')  can be seen 
\citep{DeLuca07,nlun11}. This is also the region where some of 
the highest densities, $N_{\rm e} \sim 2\EE{3} \cm3$, in the PWN can be 
found \citep{san13}. In addition to the ACA and ATCA data, we have
also used IR data collected by the AKARI satellite in 2006 for the 
wavelength range $2.53-28.7$~$\mu$m, to add to the Spitzer 
data.

The paper is organized as follows: in Section~\ref{sec:Observations} we describe and discuss the
ATCA, ACA and AKARI observations, and in Section~\ref{sec:Results} we analyse the
observations. In Section~\ref{sec:discuss} we put the results into context, and in Section~\ref{sec:Conclusions} 
we summarize our conclusions.

\begin{table}
%\centering
\caption{ACA and AKARI observations of \snr.} 
\label{tab:observations}
\begin{tabular}{lccc} % four columns, alignment for each
\hline
Time                       & Band & Integration & Frequency          \\
UT                         &           &      s           & GHz       \\      
\hline
2018 Mar 19.96 &    ACA/Band 4    &  438.48      & $138^{\rm a}$      \\
                               &          &                   & $140^{\rm a}$           \\
                               &          &                   & $150^{\rm a}$           \\
                               &          &                   & $152^{\rm a}$           \\
2018 Mar 25.94  &    ACA/Band 6    &  665.28      & $214.5^{\rm a}$       \\
                               &          &                   & $216.5^{\rm a}$     \\
                               &          &                   & $230.5^{\rm a}$           \\
                               &          &                   & $232.5^{\rm a}$           \\
2006 Oct 31.83 & AKARI/L24 & 148.75$^{\rm b}$ & $(1.044-1.629)\times10^{4}$                   \\ 
2006 Oct 31.21 & AKARI/L15 & 148.75$^{\rm b}$ &  $(1.396-2.456)\times10^{4}$                    \\
2006 Oct 25.98 & AKARI/S11 & 148.75$^{\rm b}$ & $(1.960-3.625)\times10^{4}$                     \\
2006 Oct 25.97 & AKARI/S7  & 148.75$^{\rm b}$ &  $(3.429-5.423)\times10^{4}$                    \\
2006 Oct 25.98 & AKARI/N3  & 147.0$^{\rm b}$ &  $(7.519-11.82)\times10^{4}$                    \\
\hline
\end{tabular}
$^{\rm a}$Average frequency of each spectral window (spw). The spw bandwidth is 1.875 GHz.\\
$^{\rm b}$Net integration times including both long and short exposures in three dithered sky positions.\\
%References: $^1$\citet{bra14}, $^2$This paper, $^3$\citet{Williams08}, $^4$\citet{Mignani12}, $^5$\citet{serf04}.\\
\end{table}

%  (1) Some information on how you reduced the ATCA data is needed in Section 2.1.  Did you use MIRIAD?  If so, what version?  What weighting did you use in making these images?  Did you self-calibrate the data? I believe this information is critical for others to be able to reproduce your results if they so choose, which is why I request it for all of the datasets presented in this work. 
% (2) What is the RMS sigma in the images of Figure 1?  I believe it is mentioned in the text, but should be mentioned in the Figure caption as well.
´%(3) In Figure 5, is the middle panel "Image_4shell" and the right panel "Image_4diff" as described in Equations (1) and (2).  If so, it would be good to mention this in the caption.
% (4) You need some explanation why the flux of [OIII],4959 is 1/3 of [OIII],5007.  Why not some different number, like 1/2 or 1/4?  I am not saying Williams et al. 2008 is correct, your explanation why you do something different is completely reasonable, but I feel some physical justification of why this particular value is needed.
% (5)In your discussion of the various possible breaks in the synchrotron spectrum of this PWN, it is worth mentioning / remembering that models (e.g., Gelfand et al. 2009, Bucciantini et al. 2011, Torres et al. 2014) for the spectral evolution of a PWN predict the synchrotron spectrum is not a single power law or a broken power law but a smoothly varying function.  This is because of the continuously changing injection and loss of energy from this system.
\section{Observations}
\label{sec:Observations}
\subsection{ATCA obsevations}
\label{sec:radio3cm}
\snr\ has been observed with ATCA at several frequencies on several
occasions. We have used the 3-cm data first presented in \citet{dick02}
and subsequently in \citep[][cf. their Table 1]{bra14}.
These data were obtained 1995 Oct. 23--24 UT under ATCA Program C014. Polarization was 
registered \citep[cf.][]{dick02}, but is not utilized here.
The data were reduced using \textsc{Miriad} version 
1.0\footnote{https://www.atnf.csiro.au/computing/software/miriad/} \citep{sault95}, 
setting robust = 0, and the cell-size to $0\farcs25$ to get adequate 
oversampling to reveal features with size $\lsim 1\arcsec$. 
%For the same reason 
%the parameter fwhm = 1.0 to optimize the signal-to-noise ratio in the resultant dirty image for 
%features of this size. 
The image was then cleaned using \textsc{Miriad Clean} version 1.0 with
gain=0.02 and niter=200000. Finally, restoration was done with  \textsc{Miriad Restor} version 1.2 setting
fwhm = 1.0, i.e., a circularly symmetric Gaussian beam was assumed with a full-width at half maximum of $1\farcs0$.
The final image had a
spatial resolution with a half-power
beam-width (HPBW) of 0\farcs8 (cf. Table~\ref{tab:observations3}).
The spatial resolution is thus superior to the 3-cm image discussed by
\citet{bra14}, which has a beam size of $1\farcs9 \times 1\farcs5$. We have
used our image for the analysis here, except for the integrated flux density from the PWN at 3 cm, which
we took from \citet{bra14}, as they used natural weighting for the robustness which 
trades spatial resolution for slightly better signal-to-noise.

As discussed by \citet{ser05} and \citet{nlun11} the positional
uncertainty of the pulsar, and thus the PWN, is larger in the radio ($\sim 2\arcsec$)
and in X-rays ($\sim 0\farcs5$) than in the UV/optical \citep[$\lsim 0\farcs1$,][]{Mignani19}. 
The alignment between optical and X-ray images was discussed in \citet{nlun11}, and is used
in Figure~\ref{fig:pwn_multii}. Based on the arguments in Section~\ref{sec:morph}, we have aligned
the ATCA data with the optical data in a similar way. As our analysis focuses on the PWN with its
$\sim 7\arcsec-8\arcsec$ extent, positional uncertainty is not a source of overall uncertainty 
for our analysis.

% 05 40 11.173 (0.121); −69 19 54. 41 (0. 7) [Serafimovich et al. 2005]
% 05 40 11.202 (0.009); −69 19 54. 17 (0. 05) [Mignani et al. 2019]
% Difference 0.029 seconds and 0.24 arcsec, (i.e.,  0.29 arcec)
% SWIFT catalogue 05h40m07.7s	-69d20m05s (2013)
% Kaaret et al. 2001: 05 40 11.221 ± 0.132; -69 19 54.98 ± 0.7

\begin{figure*}
\includegraphics[width=18cm]{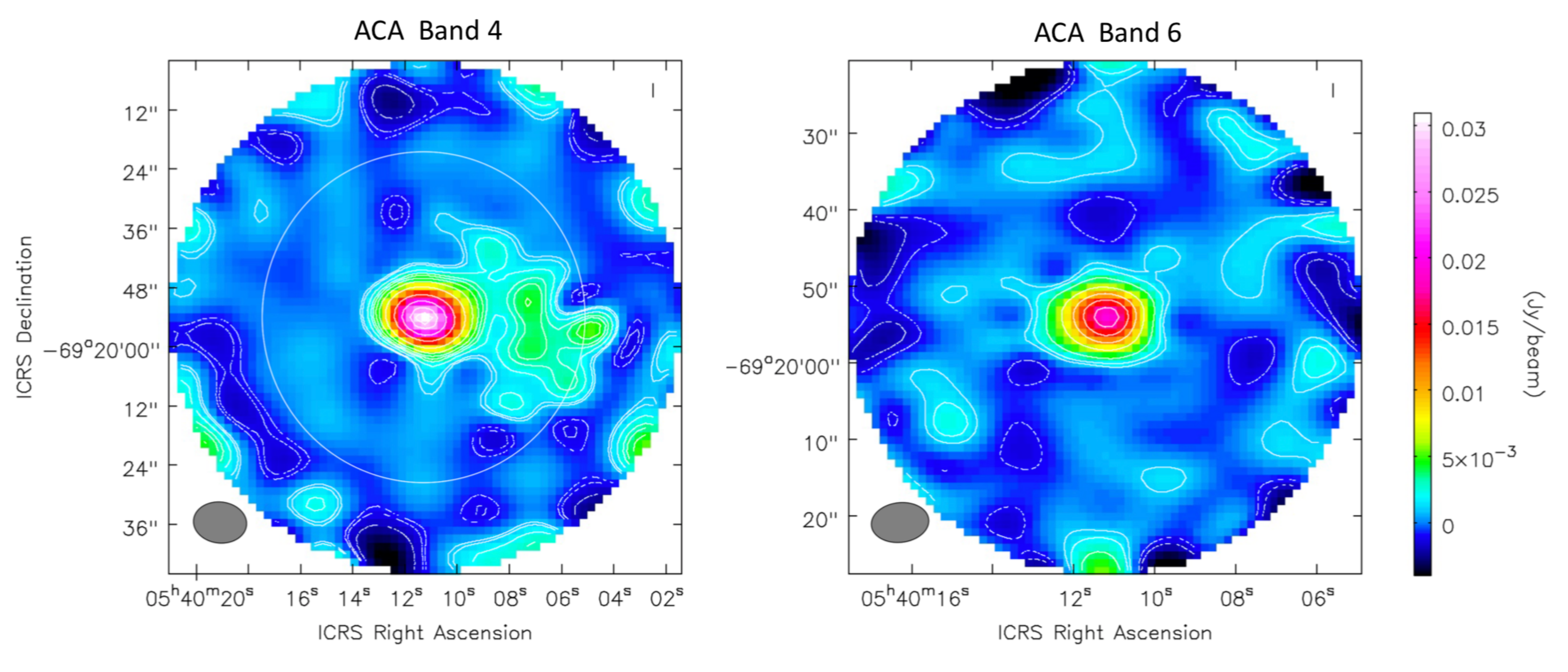}
\caption{ACA Band 4 (left panel) and Band 6 (right panel) images of \snr. The rms noise levels are 0.291~mJy and 0.348~mJy,
for Bands 4 and 6, respectively. The signal-to-noise 
(peak flux density divided by rms) for the central part of the PWN in Band 4 is $\approx 106$, and for Band 6 it is $\approx 56$. For Band 4
the western part of the remnant shell at $20\arcsec-35\arcsec$ from the center is clearly visible. 
Structures at the edges of the maps are not real. The faintest intensity shown by solid white contours marks the 
3$\sigma$ intensity level. The Band 4 contours shown are at -7, -4, -3, 3, 4, 7, 10, 12, 15, 20, 50, 70, 90, 100$\sigma$, and 
the Band 6 contours are at -5, -3, 3, 5, 10, 15, 20, 30, 40, 50$\sigma$.
The ACA Band 4 beam has size $10\farcs79 \times 8\farcs33$ and orientation 
PA~$=84\fdg09$, and the corresponding numbers for Band 6 are $7\farcs52 \times 5\farcs18$ and 
PA~$=-83\fdg60$. The beams are displayed in the lower left corners of each panel. 
The intensity scale is the same for both panels. For comparison, the Band 6 field-of-view is highlighted as a 
white circle in the Band 4 image.}
\label{fig:JB4B6}
\end{figure*}

%McMullin, J. P., Waters, B., Schiebel, D., Young, W., & Golap, K. 2007, in
%ASP Conf. Ser. 376, Astronomical Data Analysis Software and Systems
%XVI, ed. R. A. Shaw, F. Hill, & D. J. Bell (San Francisco: ASP), 127

%4            145         30.97 +/-0.29        0.291          10.79 x 8.33 (84.09)
%6            223.5      19.33 +/-0.35        0.348           7.52 x 5.18 (-83.60)
%-----------------------------------------------------------------------------------------------------------%
%Notes.
%Column (3) Peak flux density. Uncertainties are Xσ where σ is the rms noise level given in (4); values do not include 5% absolute flux calibration uncertainty. (5) Synthesized beam size (and beam position angle). (6) Integrated flux density.
\begin{table}
%\centering
\caption{ACA continuum parameters for \snr.} 
\label{tab:observations2}
\begin{tabular}{lcccc} % four columns, alignment for each
\hline
Band    & Frequency &   $S_{\nu,{\rm peak}}$$^{\rm a}$      & $\sigma_{\rm rms}$  &  $\theta_ {\rm beam}$$^{\rm b}$       \\
            &      GHz      &   mJy beam$^{-1}$  & mJy                             & size (PA)   \\      
\hline
4 &    145    &  $30.97\pm1.58$     & 0.291    &   $10\farcs79\times8\farcs33~(84\fdg09)$ \\
6  &   223.5 &  $19.33\pm1.96$     & 0.348    &   $7\farcs52\times5\farcs18~(-83\fdg60)$  \\
\hline
\end{tabular}
$^{\rm a}$Peak flux density. Uncertainties are the rms noise level given in column 4, combined with a 5\% absolute flux calibration uncertainty for Band 4, and a 10\% uncertainty for Band 6 (cf. Section~\ref{sec:fluxes}). \\
$^{\rm b}$Synthesized beam size (and beam position angle, see Section~\ref{sec:alma4a6})). \\
\end{table}

\begin{table}
%\centering
\caption{Spatial resolution of the ACA, AKARI and ATCA data for \snr.} 
\label{tab:observations3}
\begin{tabular}{lccc} % four columns, alignment for each
\hline
Instrument    & Band &    Frequency    &  Beam size     \\
                     &          &   GHz               &                         \\      
\hline
ATCA &    3 cm        &    8.64              &   $0\farcs8$$^{\rm a}$ \\
ACA &    Band 4      &    145               &   $10\farcs79\times8\farcs33$$^{\rm b}$ \\
ACA &   Band 6       &   223.5             &  $7\farcs52\times5\farcs18$$^{\rm b}$  \\
AKARI &  MIR-L L24       &   $1.30\EE4$    &  $6\farcs8$$^{\rm c}$  \\
AKARI &  MIR-L L15       &   $1.92\EE4$    &  $5\farcs7$$^{\rm c}$   \\
AKARI &  MIR-S S11       &   $2.84\EE4$    &  $4\farcs8$$^{\rm c}$   \\
AKARI &  MIR-S S7       &   $4.16\EE4$    &  $5\farcs1$$^{\rm c}$   \\
AKARI &  NIR N3       &   $9.55\EE4$    &  $4\farcs0$$^{\rm c}$   \\
\hline
\end{tabular}
$^{\rm a}$Synthesized beam size \citep[cf.][]{dick02}. \\
$^{\rm b}$Synthesized beam size (cf. Table~\ref{tab:observations2}). \\
$^{\rm c}$FWHM according to \citet{ona07}. \\

\end{table}

\subsection{ACA observations}
\label{sec:alma4a6}
We observed the millimetre/submillimetre emission in \snr\ using ACA in Band 4 (145 GHz) and Band 6 (223.5 GHz) 
as part of Atacama Large Millimeter/submillimeter Array (ALMA) program 2017.1.01391.S. The data were obtained on 
2018 March 19 and 2018 March 25 for Bands 4 and 6, 
respectively, and the array consisted of 10--11 7-m antennas. A single pointing was observed in each band. The total 
on-source integration times were 438.5 seconds and 665.3 seconds, in Bands 4 and 6, respectively.
The spectral setup covered a total usable bandwidth of 7.5 GHz, consisting of four 2-GHz (1.875 GHz usable), 
128-channel spectral windows (spws) and dual polarisation. The four spws were centred at 138, 140, 150 and 152 GHz 
for Band 4 and 214.5, 216.5, 230.5 and 232.5 GHz for Band 6. The observing parameters are summarized 
in Table~\ref{tab:observations}.

The data were reduced using the Common Astronomy Software Applications (CASA) package \citep{McM07}. We 
used the calibrated uv-data delivered by ALMA, which was calibrated using the ALMA Pipeline, but we performed our own 
imaging to improve signal-to-noise and image quality (the delivered image products were under-cleaned).
The calibrated Band 4 and Band 6 datasets, respectively, contained data from 10 and 9 antennas. We carried out 
imaging of the continuum emission, utilizing the total 7.5 GHz bandwidth, using the CASA task TCLEAN with Briggs weighting 
and a Robust = 0.5 weighting of the visibilities, using CASA 
version 5.4.0. We also performed self-calibration. For the Band 4 data, we performed several 
rounds of phase-only self-calibration, initially with a 300 second time interval followed by refinement with shorter time intervals 
down to a 30 second interval. Amplitude self-calibration was then performed with a time interval of 30 seconds. For the Band 6 
data, we performed one round each of phase-only and amplitude self-calibration, with a time interval of 300 seconds. The resulting images
 are shown in Figure~\ref{fig:JB4B6}, and the 
resulting synthesized beam sizes and rms noise levels are given in Table~\ref{tab:observations2}.  Self-calibration produced a factor 
of about two improvement in the S/N for the Band 4 data compared to deeper cleaning alone (and a factor of several compared to the 
ALMA-delivered shallow-cleaned images). For Band 6, self-calibration only produced a slight improvement (at the 20\% level) in the S/N 
compared to our deeper cleaning alone, but the appearance of imaging artifacts was improved.
%The nominal velocity 
%resolution in Bands 4 and 6 are  $61 \kms$ and $40 \kms$, respectively, but to obtain good  signal-to-noise for the continuum 
%emission, we  frequency-integrated the emission in each band.

\begin{figure*}
\includegraphics[width=17.4cm]{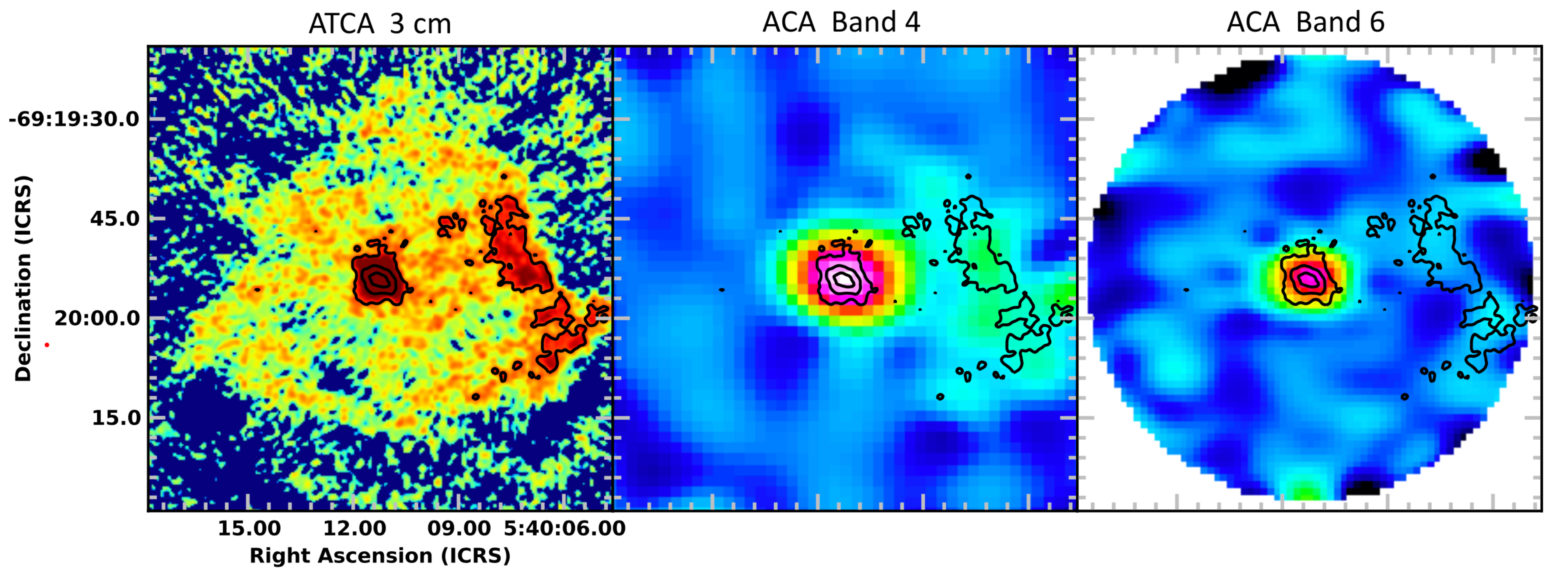}
\caption{$70\arcsec \times 70\arcsec$ maps of PWN0540 in radio (3 cm, ATCA), and at millimetre wavelengths (2 mm and 1.3 mm, 
ALMA/ACA Bands 4 and 6; middle and right panels, respectively), covering the PWN and remnant shell. The axes in all three panels 
is the same. For comparison, and to aid the eye, contours showing the 3-cm emission at levels of 0.2, 1.0 and 2.0 mJy~beam$^{-1}$ are overlaid 
on all the maps. In the 3-cm map the half-power beam-width is 0\farcs8 \citep{dick02}, and the PWN size is $\sim8\arcsec \times 7\arcsec$, so 
the PWN is fully resolved. Details for the ACA maps, including the beam sizes and intensity scale, are the same as in Figure~\ref{fig:JB4B6}. 
With the larger beams in the millimetre maps, the PWN is only partially resolved (see discussion in text). Note the similar structures of the 
western part of the remnant shell in radio and in ACA Band 4.
}
\label{fig:Bands4a6}
\end{figure*}

\subsection{AKARI observations}
\label{sec:AKARI}
\snr\ was observed in 2006 by AKARI as part of an AKARI IR
survey of the Magellanic clouds \citep{Kato12}. Five bands
covered the range $2.53-28.7$~$\mu$m, which overlaps with the
wavelength range of previously reported Spitzer observations
\citep{Williams08}. Processed data were retrieved from the
AKARI-LMC Point Source
Catalogue\footnote{www.ir.isas.ac.jp/AKARI/Archive/Catalogues/},
and entries for \pwn\ were identified for all imaging filters in the
catalogue, i.e., NIR N3 (3.2$\mu$m), MIR-S/S7 (7.0~$\mu$m), 
MIR-S/S11 (11~$\mu$m), MIR-L/L15 (15~$\mu$m) and MIR-L/L24 (24~$\mu$m). 
The observations are summarized in Table~\ref{tab:observations}, and the spatial
resolution in the various filter is listed in Table~\ref{tab:observations3}. 
The pixel scales in NIR, MIR-S and MIR-L images are 1\farcs446, 
2\farcs340 and 2\farcs384, respectively. The photometric values in the
catalogue were obtained using a radius of 10 pixels for N3, and 3 pixels
for other band images. \pwn\ was therefore fully covered by the aperture
photometry, although marginally so for the MIR-L/L24 band.

%\begin{figure}
%	\includegraphics[width=8.4cm]{Overlays.pdf}
%    \caption{ACA Band 4 map of PWN0540, with ACA Band 6 data being overlaid. Beam sizes are showed in the lower left 
%    corner, and are $11\arcsec \times 8\arcsec$ and $7\farcs5 \times 5\farcs0$, for Bands 4 and 6, respectively. Note that the
%     observed structure of the PWN in the two bands is dominated by the shape of the beam, and its orientation. For Band 6, the 
%     FWHI size is marginally smaller than the size of full PWN at 3 cm, so the PWN in this band may be marginally resolved.}
%        \label{fig:overlay}
%\end{figure}

\begin{figure*}
\includegraphics[width=17.4cm]{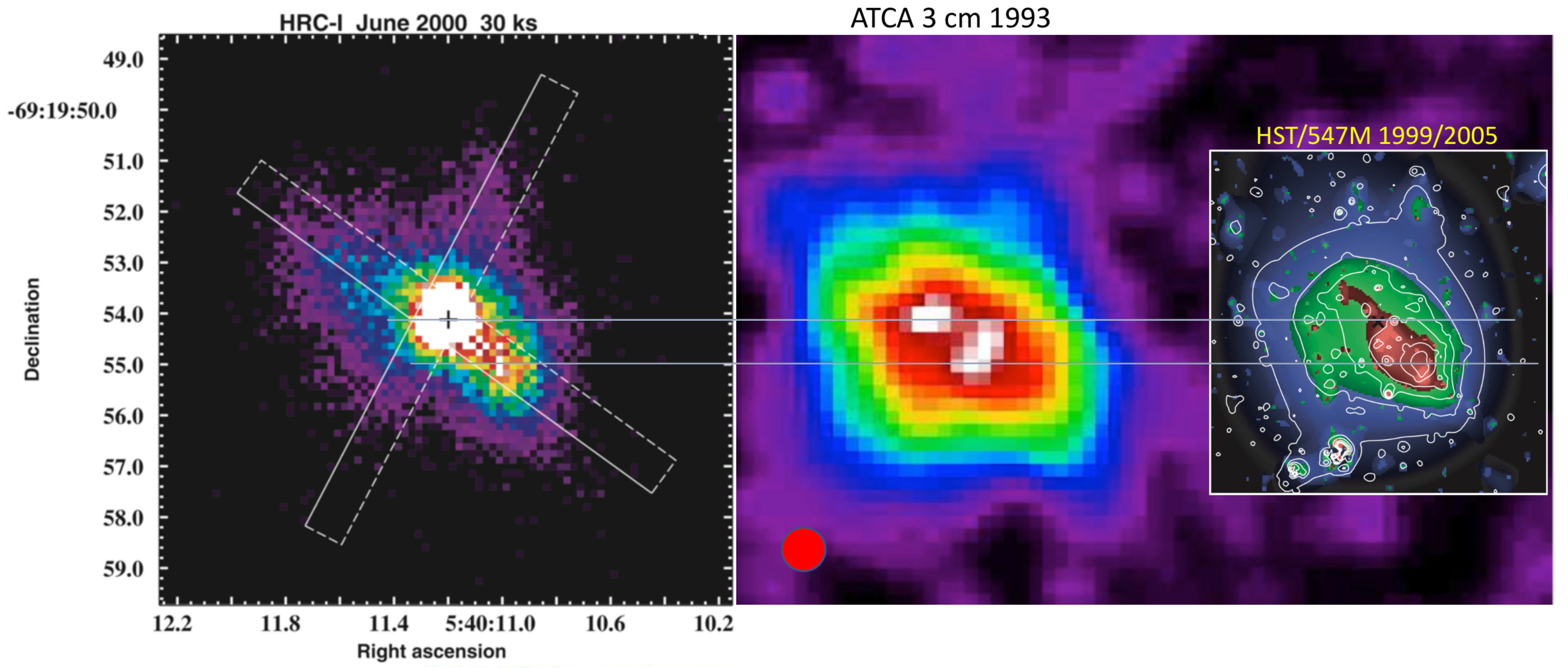}
\caption{Zoom in of the 3-cm map (middle panel), showing two regions 
of strong emission (white areas), namely the pulsar region, and the
``blob" $\sim 1\farcs5$ southwest of the pulsar. This panel is similar to 
that in \citet{dick02}, but brings out structures more clearly. The filled
red circle is the 0\farcs8 HPBW. Left panel shows a Chandra 
11\arcsec$\times$11\arcsec\ X-ray map obtained with the High-Resolution
Camera (HRC-I) in 1999. Again, the two dominating emission centres are
seen. This panel, and the rightmost panel, showing the optical continuum 
as observed with HST/F547M, are from \citet{nlun11}. Note that the 
HST colour map is for the 2005 epoch, and that contours are for 1999.
All images are to the same scale, and grey horizontal lines have been drawn through the 
pulsar position, and another through the blob at its 1999 position in the 
optical. In the optical image, data were wavelet filtered, 
after the pulsar and stars had been subtracted. The position of the 
pulsar in this panel is marked by a black cross. See text for further details.}
\label{fig:pwn_multii}
\end{figure*}

\section{Results}
\label{sec:Results}
\subsection{Morphology revealed by ACA and ATCA}
\label{sec:morph}
Figure~\ref{fig:JB4B6} displays the ACA Band 4 and 6
data. The beams and their orientations are highlighted as 
HPBW ovals in the lower left corners of the panels, and are described in 
the figure caption. The central PWN is clearly seen in both bands, as well 
as the SNR emission to the west out to $\sim 35\arcsec$ for Band 4. 
Band 6 with its smaller field-of-view and lower signal-to-noise, cannot reliably 
trace the remnant shell (see also Figure~\ref{fig:Bands4a6}).

A $70\arcsec \times 70\arcsec$ 3-cm map of \snr, using the data discussed in 
Section~\ref{sec:radio3cm}, is shown in the left panel of Figure~\ref{fig:Bands4a6}. 
Apart from the well-known strong emission on the west side, $\sim 20\arcsec-35$\arcsec\
from the pulsar, there is weaker emission in a seemingly east-west 
symmetry. In general, the outer remnant emission is more  extended on the
southern side, reaching in south-eastern direction as far as it does to 
the south-west. The extended structure on the south-eastern side has not 
been revealed in previously published radio images, although emission in
this region has been displayed in X-rays \citep{park10}. A fuller 
comparison between X-rays and radio for the SNR shell is made in
\citet{bra14}. The middle and right panels of Figure~\ref{fig:Bands4a6} 
show ACA Band 4 and Band 6 maps, respectively, with the 3-cm data being
overlaid. The larger HPBW of the ACA data smears out the remnant structures
revealed by ATCA, but there is a clear overlap of the remnant shell between radio and 
ACA Band 4. There is also a hint of overlap between radio and Band 6, but a
deeper image is needed to draw firm conclusions.

Closing in on the PWN, both the ACA bands have beam sizes larger (Band 4), 
or of roughly the same size (Band 6) as the PWN. No obvious structures of the PWN are revealed 
in the ACA bands. This is contrasted by the detailed structure of the PWN
depicted in the 3-cm radio image in the middle panel of
Figure~\ref{fig:pwn_multii}, where the radio image is drawn together with 
X-ray and optical images on the same scale. The X-ray and optical images are
from our previous study in \citet{nlun11}. The X-ray data were taken with
Chandra HRC-I in June 2000 in the $0.2-10$ keV band, and the optical HST/F547M
continuum data are from1999 and 2005, The optical data shown in
Figure~\ref{fig:pwn_multii} were wavelet-filtered by \citet{nlun11} to more
clearly bring out structures of the PWN.

\begin{figure}
\includegraphics[width=8.6cm]{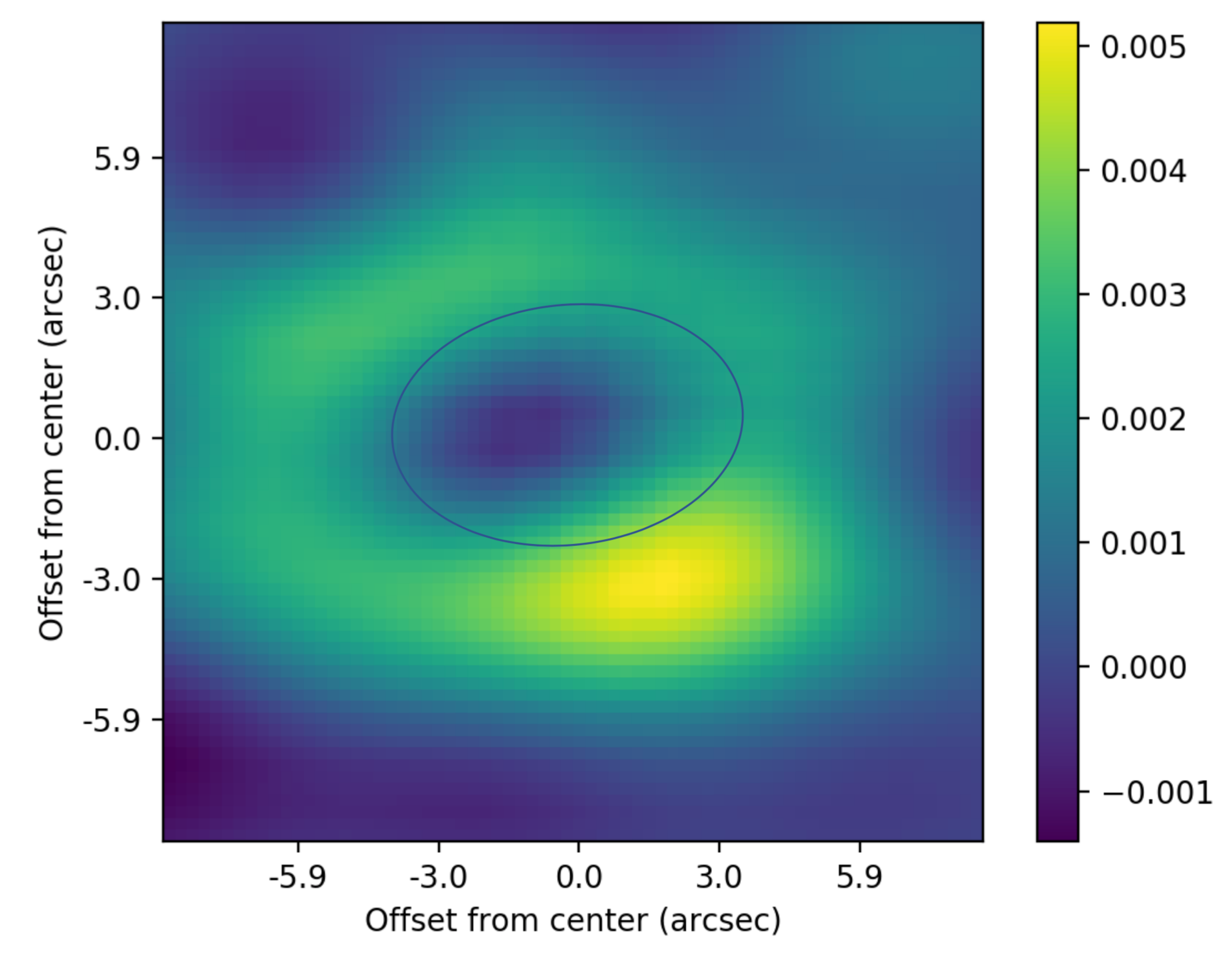}
\caption{Difference in flux density (in Jy beam$^{-1}$) between ACA Band 6 data and a point 
source smeared with the $7\farcs5 \times 5\farcs2$ beam (shown as an oval, see also
Figure~\ref{fig:JB4B6}). The modelled point source had the same surface peak
flux as in the ACA image, and was placed at the position of the peak flux density in 
the ACA image. The yellow/green ring-like structure is excess flux due to the 
extended structure of the PWN, showing that the Band 6 data are partially
spatially resolved. Note the particularly strong flux density in the direction of the
blob to the southwest (cf. Figure~\ref{fig:pwn_multii}).}
\label{fig:simulate}
\end{figure}

The 0\farcs8 HPBW of the radio data is shown Figure~\ref{fig:pwn_multii} as a
filled red circle. The spatial resolution is good enough for the 3-cm emission
to clearly reveal the two strongest continuum sources, namely the pulsar and 
its immediate surroundings, as well as the blob $\sim 1\farcs5$ southwest of the
pulsar. The two centers of emission were revealed in radio for first time by
\citet{dick02}, but there it was suggested that the pulsar was the object 
to the southwest. A comparison between the three panels of
Figure~\ref{fig:pwn_multii} clearly shows that this is not the case, and to
guide the eye, we have drawn grey horizontal lines in the east-west direction
through the position of the pulsar and through the blob at its 1999 position. 
As discussed by \citet{DeLuca07} and \citet{nlun11}, the emission centre of the
blob in the optical appears to move over time towards the southwest. This is visible in
the panel to the right of Figure~\ref{fig:pwn_multii} where the 1999 structure 
in the optical is overlaid on that from 2005. If the blob is a multi-wavelength
feature/entity, it does not come as a surprise that the blob in the X-ray image from
2000 lines up well with the optical image from the same year, and that the radio image 
from 1995 shows a blob emission centered slightly closer to the pulsar than on
images from 1999/2000. From Figure~\ref{fig:pwn_multii} this seems
to be the case. However, the overall structure of the PWN is similar at 
all wavelengths, although the relative flux densities within the PWN vary with frequency.

%\begin{figure}
%\includegraphics[width=8.2cm]{FigXX.pdf}
%\caption{Difference in flux between ACA Band 4 data and 3-cm data smeared
%with the ACA Band 4 beam. The PWN region and the rest of the remnant were allowed
%to have different spectral indices between radio and millimetre wavelengths. Note that....}
%\label{fig:subtract}
%\end{figure}

\subsection{Fluxes in the ACA and AKARI bands}
\label{sec:fluxes} 
The peak flux density of SNR 0540-69.3 in Band 4 is  30.97~mJy beam$^{-1}$ and the rms 0.291~mJy beam$^{-1}$. For Band 6 the 
corresponding numbers are 19.33~mJy beam$^{-1}$ and 0.348~mJy beam$^{-1}$, respectively. These and other image properties 
are given in Table~\ref{tab:observations2}. While the S/N of \pwn\ in Band 4 is $\approx 106$ after our improved 
cleaning and self-calibration (see Section~\ref{sec:alma4a6}), the S/N of the brightest pixel in the SNR shell features is $\sim$18, 
so it is clear that the factor of $\sim$2 improvement in S/N provided by self-calibration results in a significantly more robust 
detection of this extended feature (and we note that these SNR shell features were not detected in the initial shallow-cleaned images; 
the deeper cleaning and self-calibration described in Section~\ref{sec:alma4a6} were essential for the detection of these features). 
Regarding uncertainties, as stated in Table~\ref{tab:observations2}, there is an uncertainty in the absolute flux calibration, 
which is estimated to be  5\% for Band 4 and 10\% for Band 6, as detailed in the ALMA Proposer’s Guide\footnote{https://almascience.eso.org/documents-and-tools/cycle7/alma-proposers-guide} (Section A.9.2) 
and the ALMA Technical Handbook\footnote{https://almascience.eso.org/documents-and-tools/latest/documents-and-tools/cycle7/alma-technical-handbook} (Chapter 10). The flux uncertainty quoted in Table~\ref{tab:observations2} is a combination of this and the rms.

%To bring out the most from the delivered Pipeline-calibrated ACA data, we
%improved the imaging stage of the reductions as described
%in Section~\ref{sec:alma4a6}. In the Pipeline-calibrated image, the signal-to-noise (S/N) 
%for the peak flux was $\sim 14$ and $\sim 10$ for Bands 4 and 6, respectively. After the improved 
%data calibration and cleaning, the peak flux for the spatial distribution in Band 4 is 30.97~mJy beam$^{-1}$ and the 
%rms 0.291~mJy beam$^{-1}$. For Band 6 the corresponding numbers are 19.33~mJy beam$^{-1}$ 
%and 0.348~mJy beam$^{-1}$, respectively. This means that our improved data calibration and cleaning 
%allowed us to detect $6-7$ times fainter features than in the delivered data. We would, e.g., not have
%been able to identify and study the SNR shell in Band 4, had we not performed self-calbration.
%The brightest SNR shell features in our images have S/N~$\sim 15$, whereas they have S/N~$< 3$ in the delivered data.

%As stated in Table~\ref{tab:observations2} there
%is an uncertainty in the absolute flux calibration, which is estimated to be $\sim 5\%$ for Band 4
%and 10\% for Band 6, as detailed in the ALMA Proposer’s Guide\footnote{https://almascience.eso.org/documents-and-tools/cycle7/alma-proposers-guide} (Section A.9.2) and the ALMA Technical 
%Handbook\footnote{https://almascience.eso.org/documents-and-tools/latest/documents-and-tools/cycle7/alma-technical-handbook} (Chapter 10). The flux uncertainty quoted in Table~\ref{tab:observations2} is a combination of this and the rms.

The integrated flux density from the PWN would be the same as the peak flux density if the PWN 
were unresolved. This is not the case, since \pwn\ is partially resolved in both ACA bands. To test this we made
an experiment were we assume that the source is a point source, so that the
modeled spatial flux density distribution essentially becomes the beam distribution. 
We then normalized the peak flux density of this image to the observed image for each band. 
The difference between the observed image and the model is shown in
Figure~\ref{fig:simulate} for Band 6. The net flux density is at zero level in the 
center, as expected, but positive residuals are clearly displayed outside the 
HPBW oval, especially in the direction towards the blob. 

We can use the result from this experiment to estimate a correction factor
for the integrated flux density from the PWN, compared to the observed peak flux densities. 
This correction factor is simply the net flux density, as displayed in Figure~\ref{fig:simulate} for Band 6, 
relative to that measured from a point source. We obtain the correction factors 
$\sim1.19$ and $\sim1.50$ for Bands 4 and 6, respectively. The correction factor,
and its estimated uncertainty, are included in Table~\ref{tab:fluxtable} for Band 6.
The situation is more complicated for Band 4, since for this band the correction factor
is affected by flux from the SNR shell being smeared into the PWN region.
 
For Band 4 we have therefore also made another test which relies on an assumption that 
the underlying structures of the PWN and the SNR shell are similar at 3 cm and in the ACA bands. 
This assumption is best checked for Band 6 image with its higher spatial resolution, by smearing 
the 3 cm image with the ACA Band 6 beam, and then subtract this from the observed ACA Band 6 
image. The residuals are found to be consistent with noise, which means that the intrinsic structures are 
similar at 3 cm and in Band 6 (and therefore also in Band 4), at least at the levels of signal-to-noise and 
spatial resolution of the ACA bands.  

\begin{figure*}
\includegraphics[width=17.9cm]{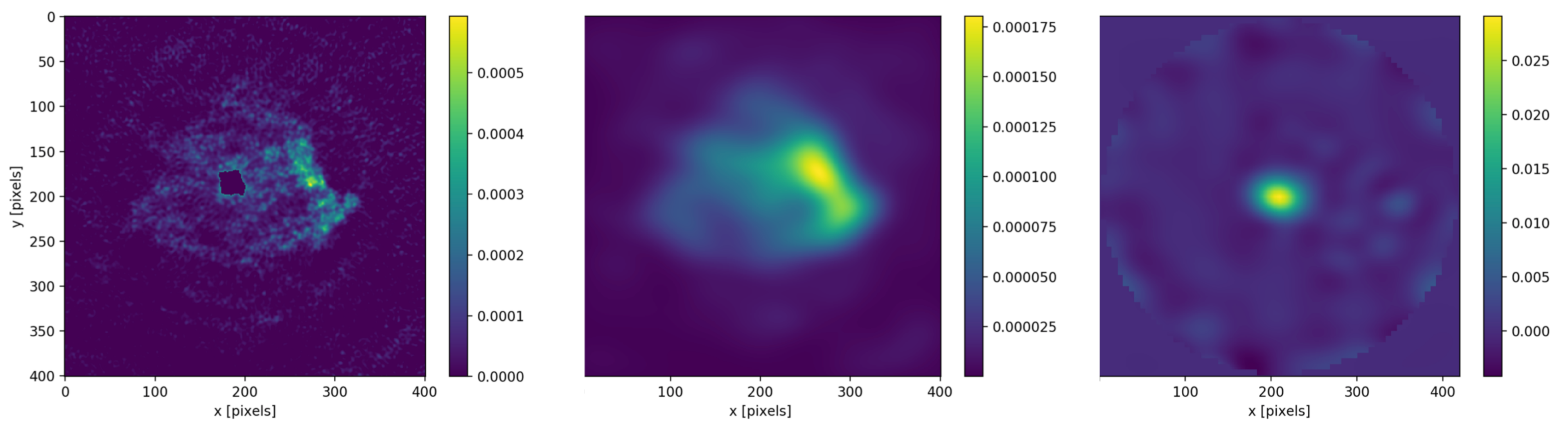}
\caption{3-cm image of \snr, with the PWN masked out (left panel). To obtain the middle panel we smeared
the 3-cm image with the ACA Band 4 beam. This simulates the structure of the pure remnant shell in Band 4, and is
named Image$_{4{\rm shell}}$ in Section~\ref{sec:fluxes}. The 
image in the panel to the right is the observed ACA Band 4 shown in the left panel of Figure~\ref{fig:JB4B6} (and called 
Image$_{4{\rm obs}}$ in Section~\ref{sec:fluxes}) minus Image$_{4{\rm shell}}$. The fux densities are in Jy beam$^{-1}$. 
%See text for further details.
}
\label{fig:Band4}
\end{figure*}

This intrinsic similarity between radio and millimetre structures allows us to improve on the correction 
for the integrated flux density of the PWN in Band 4. We did this by first constructing two images. For the
first image we only considered the PWN area of the 3-cm map, and smeared this 
with the ACA Band 4 beam. We call this Image$_{4{\rm PWN}}$. We also made a simulated image 
of the SNR shell, Image$_{4{\rm shell}}$, where we first masked out the PWN from the 3-cm map, and then
smeared the rest with the ACA Band 4 beam. We show this process in Figure~\ref{fig:Band4}. We then removed 
the simulated shell image Image$_{4{\rm shell}}$ from the observed Band 4 image, Image$_{4{\rm obs}}$, 
thereby creating the right panel of Figure~\ref{fig:Band4}. This last step can be written
\begin{equation}
    \text {Image}_{4\text {diff}} =  \text {Image}_{4\text {obs}} - B \times \text {Image}_{4\text {SNR}}~.
\label{eq:images1}
\end{equation}
We then created a null image by removing Image$_{4{\rm PWN}}$ from Image$_{4{\rm diff}}$, i.e.,
\begin{equation}
    \text {Image}_{4\text {resid}} =  \text {Image}_{4\text {diff}} - A \times \text {Image}_{4\text {PWN}}~.
\label{eq:images2}
\end{equation}
The difference was tuned through the constants $A$ and $B$ to achieve zero net flux density at the positions of 
the PWN and SNR, respectively.
The ratio $A/B$ can be used to estimate a power-law index for the spectral range between 3 cm and ACA Band 4
for the SNR shell emission, if the index is known for the PWN.
\begin{equation}
    \alpha_{\nu,\text {SNR}} = \alpha_{\nu,\text {PWN}} + 0.816~\text{lg}(A/B).
\label{eq:index}
\end{equation}
In our analysis we arrive at $A/B = 3.8\pm0.2$, so that $\alpha_{\nu,\text {SNR}} - \alpha_{\nu,\text {PWN}} = 0.47\pm0.03$.
\citet{bra14} estimated $\alpha_{\nu,\text {SNR}} - \alpha_{\nu,\text {PWN}} = 0.50\pm0.04$ for radio alone. 
Our result is fully consistent with the same difference in power-law index continuing into the millimetre regime.

The peak flux density of the image in the right panel of Figure~\ref{fig:Band4}, i.e., Image$_{4{\rm diff}}$, is 29.07~mJy beam$^{-1}$, 
which is  $\sim 6 \%$ lower than before the correction for emission from the remnant shell (cf. Table~\ref{tab:observations2}). 
Finally, we used Image$_{4{\rm diff}}$ to calculate the flux correction factor for the PWN, as we did for Band 6, to arrive
at the value $1.05$, which we include in the integrated flux density for Band 4 in Table~\ref{tab:fluxtable}. 
This correction factor is smaller than the value we obtained before compensating for the SNR shell, as expected.

For the AKARI data, we used 343.34, 74.956, 38.258, 16.034, 
and 8.0459 Jy as fluxes for zero magnitude for the five IRC
bands N3, S7 S11, L15 and L24, respectively 
\citep{tan08,Kato12}. The AKARI flux densities are listed in
Table~\ref{tab:fluxtable}, along with the frequency interval
for which the normalized filter transparency is $\geq 50\%$. 

% Flux table
\begin{table}
%\centering
\caption{Observed multi-wavelength integrated flux densities of \pwn.$^{\rm a}$} 
\label{tab:fluxtable}
\begin{tabular}{lccr} % four columns, alignment for each
\hline
Frequency                        & Instrument & Flux density & Source$^{\rm b}$ \\
GHz                                    &                   & mJy  &                  \\
\hline
$1.513\pm0.064$   & ATCA 20~cm & $68\pm7$$^{\rm c}$   &  1\\
$2.290\pm0.064$   & ATCA 13~cm& $62\pm6$$^{\rm c}$  &  1\\
$4.790\pm0.064$   & ATCA 6~cm & $54\pm5$$^{\rm c}$  &  1\\
$5.824\pm0.064$   & ATCA 6~cm & $60\pm6$$^{\rm c}$  &  1\\
$8.640\pm0.064$   & ATCA 3~cm & $51\pm5$$^{\rm c}$  &  1\\
$145.0\pm7.9$   & ACA Band 4 & $30.5\pm1.6$$^{\rm c,d}$  &  2\\
$223.5\pm9.9$   & ACA Band 6 & $29.1\pm3.0$$^{\rm c}$  &  2\\
$4350\pm110$   & Spitzer MIPS Ch2 & $<366$  &  3\\
$(1.26\pm0.21)\EE{4}$   & Spitzer MIPS Ch1 & $13.19\pm3.95$  &  3\\
$(1.30\pm0.15)\EE{4}$   & AKARI MIR-L L24 & $17.9\pm1.2$  &  2\\
$(1.92\pm0.36)\EE{4}$   & AKARI MIR-L L15 & $6.29\pm1.02$$^{\rm c}$  &  2\\
$(2.84\pm0.54)\EE{4}$   & AKARI MIR-S S11 & $4.62\pm0.27$$^{\rm c}$  &  2\\
$(3.94\pm0.73)\EE{4}$   & Spitzer IRAC Ch4 & $5.10\pm0.74$$^{\rm c}$  &  3\\
$(4.16\pm0.50)\EE{4}$   & AKARI MIR-S S7 & $3.38\pm0.24$$^{\rm c}$  &  2\\
$(5.31\pm0.66)\EE{4}$   & Spitzer IRAC Ch3 & $3.61\pm0.46$$^{\rm c}$  &  3\\
$(6.76\pm0.76)\EE{4}$   & Spitzer IRAC Ch2 & $2.19\pm0.27$$^{\rm c}$  &  3\\
$(8.54\pm0.90)\EE{4}$   & Spitzer IRAC Ch1 & $1.77\pm0.23$$^{\rm c}$  &  3\\
$(9.55\pm1.70)\EE{4}$   & AKARI NIR N3 & $2.26\pm0.20$$^{\rm c}$  &  2\\
$(1.41\pm0.12)\EE{5}$   & VLT/NACO/$K_s$ & $0.801\pm0.035$$^{\rm c}$  &  4\\
$(1.83\pm0.19)\EE{5}$   & VLT/NACO/$H$& $0.675\pm0.027$$^{\rm c}$  &  4\\
$(2.36\pm0.20)\EE{5}$   & VLT/NACO/$J$& $0.541\pm0.022$$^{\rm c}$  &  4\\
$(3.68\pm0.57)\EE{5}$   & HST/F814W$^{\rm e}$ & $0.481\pm0.014$  &  4\\  %7061-9612 (50%). (Effective 8353)
$(3.76\pm0.43)\EE{5}$   & HST/F791W & $0.437\pm0.009$$^{\rm c}$  &  5\\ %7177-9001 (50%). (Effective 8095)
$(3.76\pm0.43)\EE{5}$   & HST/F791W & $0.426\pm0.013$$^{\rm c}$  &  4\\ %7177-9001 (50%). (Effective 8095)
$(4.48\pm0.43)\EE{5}$   & HST/F675W$^{\rm f}$ & $0.472\pm0.018$  &  4\\ %6111-7401(50%). (Effective 6727)
$(5.50\pm0.36)\EE{5}$   & HST/F547M & $0.267\pm0.005$$^{\rm c}$  &  5\\ %5121-5828 (50%). (Effective 5475)
$(5.71\pm0.86)\EE{5}$   & HST/F555W$^{\rm g}$ & $0.346\pm0.017$  &  4\\  %4572-6174 (50%). (Effective 5412)
$(6.89\pm1.13)\EE{5}$   & HST/F450W$^{\rm h}$ & $0.361\pm0.022$  &  4\\  %3738-5200 (50%) (Effective 4480)
$(9.00\pm0.64)\EE{5}$   & HST/F336W & $0.124\pm0.002$  &  5\\ %3111-3583 (50%) (Effective 3325)
$(9.00\pm0.64)\EE{5}$  & HST/F336W & $0.211\pm0.022$$^{\rm c}$  &  4\\  %3111-3583 (50%) (Effective 3325)
\hline
\end{tabular}
$^{\rm a}$Near-infrared observations, and observations at higher frequencies, are dereddened integrated flux densities, according to their sources.\\
$^{\rm b}$References: (1) \citet{bra14}, (2) This paper, (3) \citet{Williams08}, (4) \citet{Mignani12}, (5) \citet{serf04}.\\
$^{\rm c}$Included in the power-law fits in Figures~\ref{fig:multi} and~\ref{fig:multi2}\\
$^{\rm d}$Includes correction for extended emission from the SNR shell.\\
$^{\rm e}$Includes [\ion{S}{III}]~$\lambda\lambda9069,9532$.\\
$^{\rm f}$Includes [\ion{N}{II}]~$\lambda\lambda6548,6583$, H$\alpha$ and [\ion{S}{II}]~$\lambda\lambda6716,6731$.\\
$^{\rm g}$Includes [\ion{O}{III}]~$\lambda\lambda4959,5007$.\\
$^{\rm h}$Includes [\ion{O}{II}]~$\lambda\lambda3726,3729$ and  [\ion{O}{III}]~$\lambda\lambda4959,5007$.
\end{table}

%Power-law parameters  -0.87455E+00  0.98395E-01 -0.78806E-01
\section{Discussion}
\label{sec:discuss}
\begin{figure}
\includegraphics[width=8.5cm]{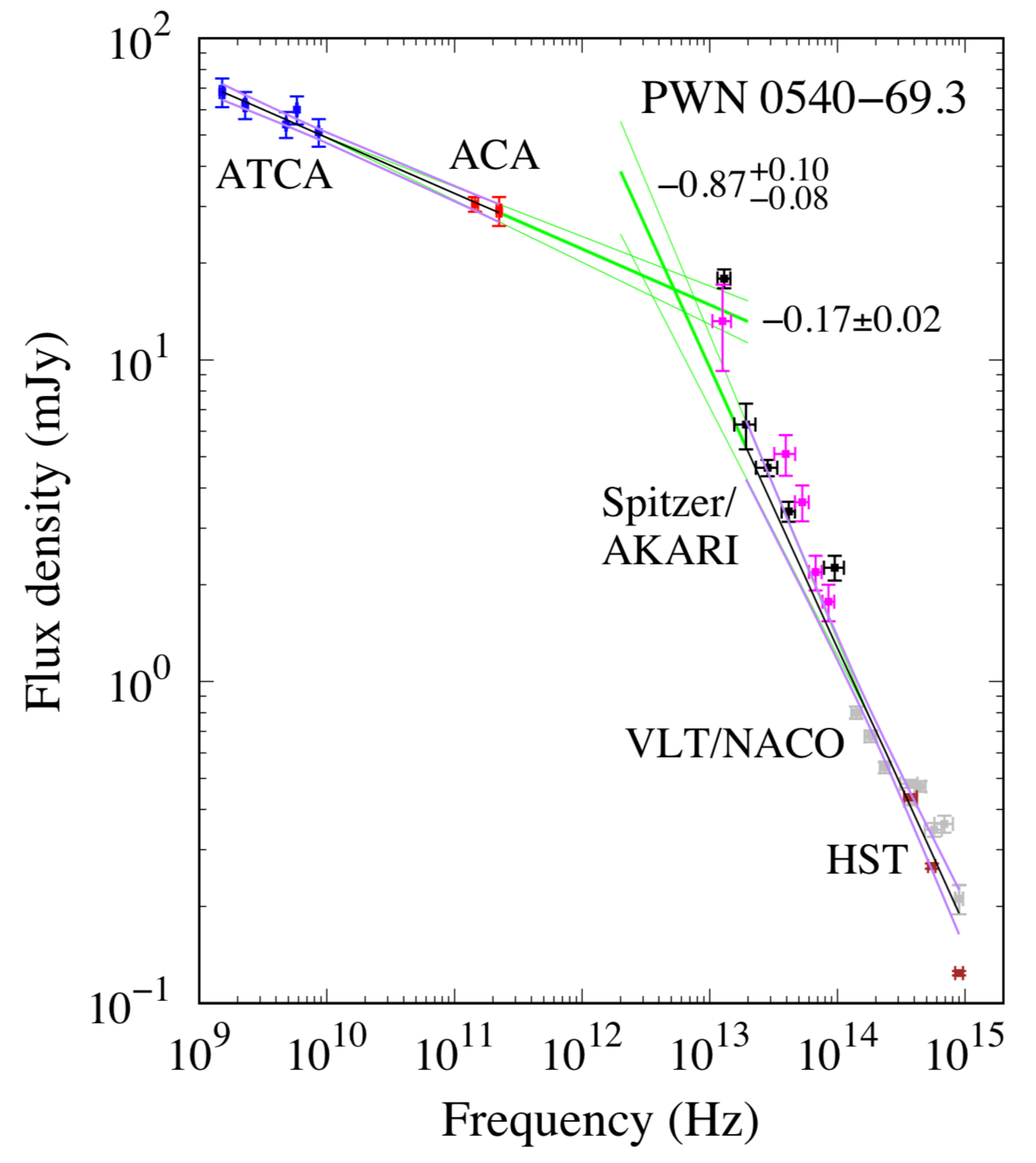}
\caption{Multi-wavelength spectrum of PWN0540. Our estimated flux densities in 
ACA Bands 4 and 6 are shown by red symbols, and the radio flux densities estimated by
\citet{bra14} are marked in blue. The best power-law fit to those data has a 
spectral index of $\alpha_{\nu,{\rm mm}}= 0.17\pm0.02$. (The figure gives
spectral slopes, i.e., $-\alpha_{\nu}$.) A similar fit to the UV/IR part gives 
$\alpha_{\nu,{\rm UVIR}}= 0.87^{+0.08}_{-0.10}$. 1$\sigma$ intervals for the power-laws 
are drawn in purple. The data included in this fit are the Spitzer
IRAC data of \citet{Williams08} (in magenta), AKARI data reported here (in black),
and the VLT/NACO data of \citet{Mignani12} (in grey). Data for the UV/optical 
regime published  by \citet{serf04} (in brown) and by \citet{Mignani12} (in grey)
are all shown in the figure, but as detailed in the text, not all those data were
included in the power-law fit. Extrapolations of the two power-laws (in green),
intersect at $\nu \sim 5.2\times10^{12}$ Hz. Extrapolations of the 1$\sigma$
intervals are also drawn in green.}
\label{fig:multi}
\end{figure}

We have used integrated flux densities in Table~\ref{tab:fluxtable} to fit power-laws
to different parts of the synchrotron spectrum. Not all data in the table have
been included in these fits, as will be detailed below. Concentrating first on the
radio/mm-region, we have used the ACA flux densities in Section~\ref{sec:fluxes}
together with radio flux densities from \citet{bra14}. We note that there is some
inconsistency in \citet{bra14} with regard to the 6-cm flux densities at 4.790 GHz and
5.824 GHz in their Figures~6 and 11, compared to their Table 2. We have assumed
that the flux densities in their Table 2 are correct (since they agree with their
Figure~11), and have used the flux densities uncertainties shown in their Figure~6 (as no
uncertainties are provided in their Table 2). With this caveat in mind, we
estimate $\alpha_{\nu} = 0.17\pm0.02$ for the frequency interval 
$1.449 - 233.50$~GHz. This is shown in Figure~\ref{fig:multi}. 

The uncertainty of $\alpha_{\nu,{\rm radio/mm}}$ and its 1$\sigma$ interval were estimated 
using the same method as in \citet{serf04}, i.e., utilizing a Monte Carlo 
approach to construct 10\,000 power-laws to fit 10\,000 simulated sets of data,
where the flux uncertainty is assumed to have a Gaussian distribution, and the
frequency distribution to be evenly distributed within the frequency bins given 
in Table~\ref{tab:fluxtable}. We have then ordered the 10\,000 simulated data 
sets in order of fitted $\alpha_{\nu,{\rm radio/mm}}$ values, and from this assigned the median
value to be the preferred $\alpha_{\nu,{\rm radio/mm}}$. We did this 500
times with different seed values for the randomizer, and then took the average
value for the power-law index to be the final estimate of the index. The 
1$\sigma$ boundaries for $\alpha_{\nu,{\rm radio/mm}}$ shown in Figure~\ref{fig:multi} come 
from the constraint that 68\% of the constructed power laws must lie within 
the 1$\sigma$ boundaries. 
  
To make a similar fit to the IR/UV part of the spectrum, we included the 
Spitzer IRAC IR data from  \citet{Williams08}, the AKARI data in
Table~\ref{tab:fluxtable} and the VLT/NACO data of \citet{Mignani12}. HST data
for \pwn\ in the optical/UV have been published by \citet{serf04} and
\citet{Mignani12}. In \citet{serf04} care was taken to only consider HST filters
which do not include strong nebular spectral lines. On the contrary, most of the
filters considered by \citet{Mignani12} include the strongest lines in the nebula
\citep[cf.][]{Kirshner89,ser05,Morse06}, as we have indicated in
Table~\ref{tab:fluxtable}. To check the contribution from spectral lines to 
the flux in HST filters, we have chosen to study the influence by
[\ion{O}{III}]~$\lambda\lambda$4959,5007 on the flux in F555W. 
Fluxes for various lines from the whole PWN were estimated by \citet{Williams08}. 
For [\ion{O}{III}]~$\lambda$5007 they used the line flux deduced by
\citet{Morse06}, and assumed that the size of the PWN, as seen in [\ion{O}{III}], is a factor of
$\approx 4$ larger than the area covered by the slit used by \citet{Morse06}. This 
means that the dereddened line flux from the whole PWN is 
$f_{[\ion{O}{III}]\lambda5007} \approx 2.7\EE{-13}~{\rm erg~cm}^{-2}~{\rm s}^{-1}$.
However, the slit used by \citet{Morse06} covers the brightest areas
in [\ion{O}{III}] \citep[cf.][]{san13}, so a correction factor of 4 is too large.
We have taken a more conservative approach for the flux of 
[\ion{O}{III}]~$\lambda\lambda$4959,5007 than \citet{Williams08}, and adopt
$f_{[\ion{O}{III}]\lambda5007} = 2.0\EE{-13}~{\rm erg~cm}^{-2}~{\rm s}^{-1}$
and $f_{[\ion{O}{III}]\lambda4959}$ to be one third of that, in accordance 
with the 3:1 ratio of the transition probabilities of the two lines \cite[e.g.,][]{ost06}. To compare 
with the continuum emission, we use the flux in the F547M filter 
(cf. Table~\ref{tab:fluxtable}) and the continuum spectral slope
$F_{\nu} \propto {\nu}^{-0.87}$ (see below). With the filter transmission 
considered, we arrive at an [\ion{O}{III}]~$\lambda\lambda$4959,5007 flux which 
is $\sim 60$\% of the continuum flux within F555W, i.e, the integrated continuum 
flux density at $5.71\times10^{14}$~Hz, is probably closer to $0.20-0.25$~mJy
than the $\sim 0.35$~mJy listed in Table~\ref{tab:fluxtable}. Had we used
the [\ion{O}{III}]~$\lambda$5007 line flux estimated by \citet{Williams08}, the derived
ntegrated continuum flux density in F555W would have been $\lsim 0.20$~mJy. Based on this, 
we have discarded HST filters which encapsulate strong nebular lines from our
analysis.

The HST filters we have included are F336W, F547M and F791W. For
F791W, the fluxes estimated by \citet{serf04} and \citet{Mignani12}, using the same
data from 1999 October 17, agree to within 
3\%, and we used an average value of those. This contrasts the situation for F336W
where the two teams obtained very different results. \citet{Mignani10,Mignani12} argue
that this is not mainly because the groups analyze different data sets (from 1999 October 17
and 2007 June 21), but that the so-called Charge Transfer Efficiency (CTE) effect, which is
important for F336W, is treated better by them. We have therefore chosen to only
include the \citet{Mignani12} measurement for that filter.
%We have chosen the flux with 
%the largest error for this filter, i.e., the one of \citet{Mignani12}, not to overly
%influence the power-law fit. 
F547M does not include any strong nebular lines
\citep{serf04}, and this band was included in our power-law estimate. All data used
in our power-law fits are highlighted in Table~\ref{tab:fluxtable} by the asterisk
marked ``c''. With this in mind, we arrive at $\alpha_{\nu,{\rm UVIR}}= 0.87^{+0.08}_{-0.10}$
for the IR/UV part of the synchrotron spectrum. 
\citep[Had we included the F336W flux of][the power-law index would only have changed to $0.91^{+0.08}_{-0.10}$.]{serf04} 
This is less steep than 
$\alpha_{\nu} = 1.48^{+0.09}_{-0.08}$ by \citet{serf04} and $1.27\pm0.3$ by
\citet{bra14}, but steeper than  $\alpha_{\nu} = 0.56\pm0.03$ by
\citet{Mignani12}. Our results supersede all those estimates. The
$-0.87^{+0.10}_{-0.08}$ spectral slope for the IR/UV range is
drawn and marked in Figure~\ref{fig:multi}. Note the flux excess for the 
discarded HST filters compared to the power-law fit, due to line emission
contamination. 

In the IR, the integrated AKARI flux densities in the range
$(2.3-4.7)\times10^{13}$~Hz are lower than those measured
by Spitzer, despite the fact that there is line contribution mainly from to
[\ion{S}{IV}]~10.5~$\mu$m \citep[][see also Figure~\ref{fig:S11}]{Williams08}. 
This underlines difficulties in 
background subtraction in the IR. In general, the AKARI data line up better with
the full IR/UV power-law, except for the N3 and L24 bands. The L24 band is
not part of the power-law fits in Figure~\ref{fig:multi} though.

In Figure~\ref{fig:multi}, an extrapolation of the high-energy power-law 
undershoots the flux level of the $\sim 24~\mu$m AKARI data by several $\sigma$. This 
could signal a contribution to the emission from dust with 
$T_{\rm dust} \sim 40-65$~K, as argued for by \citet{Williams08}. However, 
broad-line emission in [\ion{C}{IV}]~25.9~$\mu$m and [\ion{Fe}{II}]~26.0~$\mu$m
contributes to the integrated AKARI flux density in AKARI L24 band. Using the line fluxes
estimated by \citet{Williams08}, and taking into account the 
transparency in the L24 band at 26~$\mu$m\footnote{http://svo2.cab.inta-csic.es/svo/theory/fps3/index.php?mode=browse}, 
we estimate that lines can contribute up to $\sim5-6$~\% of the 
integrated AKARI flux density in the L24 band, which only marginally decreases the need for dust 
to fit the $24~\mu$m data (see below).

Extrapolations of the low-frequency and the IR/UV power-laws in
Figure~\ref{fig:multi} intersect at $\nu_{\rm break} \sim 5.2\times10^{12}$ Hz. 
If we use the 1$\sigma$ limits in Figure~\ref{fig:multi} we obtain 
$\nu_{\rm break} = (5.2^{+3.6}_{-2.8})\times10^{12}$~Hz. If one assumes that
this break is due to synchrotron cooling of relativistic electrons, one can estimate the magnetic
field strength, $B$, where the electrons are being injected. Equalling the life-time of
synchrotron-emitting electrons, $\tau_{\rm synch} = 6\times10^{11} B^{-3/2} \nu_{\rm break}^{-1/2}$~s
\citep{pac70,bra14}, to the age of the remnant, assumed to be
1\,100 years \citep{rey85}, this translates into 
$B = (3.9^{+1.1}_{-0.7}) \times10^{-4}$~G. This is higher than the 
findings of \citet{man93b} and \citet{bra14}, who both estimated 
$B \sim 2.5\times10^{-4}$~G, but somewhat lower than a recent value derived from X-rays,
$B = (7.8^{+45}_{-2.8})\times10^{-4}$~G \citep{Ge19}. 

\begin{figure}
\includegraphics[width=8.7cm]{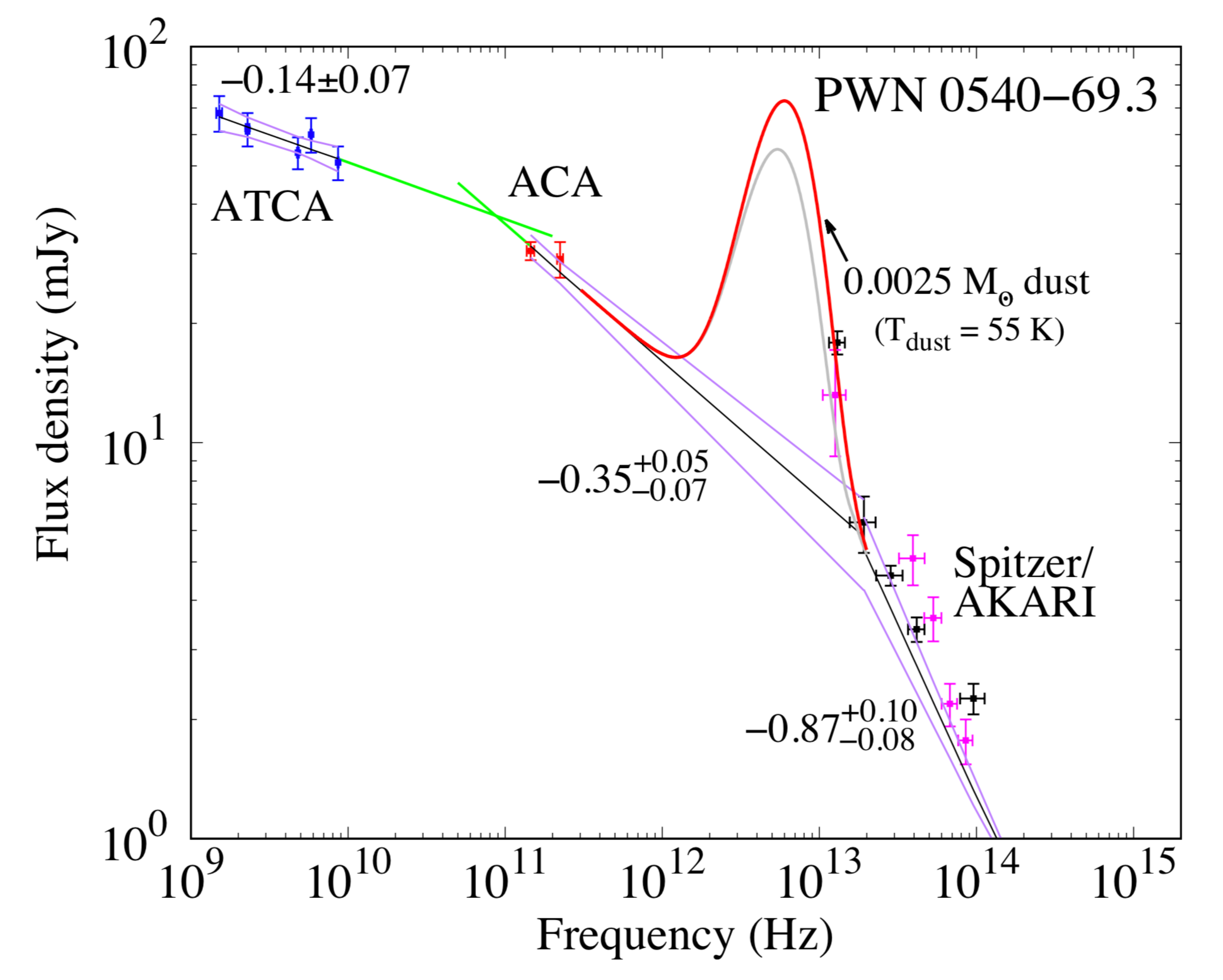}
\caption{Same as Figure~\ref{fig:multi}, but only showing the ATCA, ACA, 
Spitzer and AKARI data. One separate power-law fit has been made for the ATCA
data, and another for the ACA data plus the AKARI 15~$\mu$m data. Spectral
indices for the various frequency intervals are shown 
($\alpha_{\nu,{\rm radio}} = 0.14\pm0.07$ and 
$\alpha_{\nu,{\rm mm/IR}} = 0.35^{+0.07}_{-0.05}$). (The figure gives
spectral slopes, i.e., $-\alpha_{\nu}$.) Extrapolations of the
radio--IR power-laws intersect at $\sim 8.4\EE{10}$~Hz, which can compared with a
similar break at $\sim 3\EE{10}$~Hz break for the Crab Nebula. Emission from
$2.5\times10^{-3} \Msun$ of
silicate dust with temperature $T_{\rm dust} = 55$~K is also included (in red), 
to model the AKARI and Spitzer data long-ward of $\sim 20~\mu$m. For a comparison
we also include a model (in grey) with preferred values by \citet{Williams08}, namely the dust mass
$3\times10^{-3} \Msun$ and $T_{\rm dust} = 50$~K.}
\label{fig:multi2}
\end{figure}

The spectral break between the low- and high frequency parts of the synchrotron
spectrum gives a steepening in power-law index by $\Delta = 0.70^{+0.08}_{-0.10}$,
which is slightly larger than the expected standard change of 0.5 due to synchrotron
losses in a homogeneous medium. Although $\Delta = 0.5$ is not fully excluded from our
spectral fits, \citet{rey09} discusses possibilities with $\Delta > 0.5$, and 
how this could signal inhomogeneities in regions where the relativistic electrons
are injected. There is also the possibility that the $\Delta \sim 0.7$ break in power-law
index is not due to synchrotron losses, but signals two populations of
relativistic electrons, as in the model of \citet[][cf. Introduction]{bucc11}.

Rather than a single power-law break at 
$\nu_{\rm break} \sim 5.2\times10^{12}$~Hz, a dual break is also possible. 
The Crab PWN may serve as a template. As described in the
Introduction, the Crab PWN has a spectral break at $\sim 3\EE{10}$~Hz,
where the spectral index changes from $\alpha_{\nu} \approx 0.3$ in the radio to
$\alpha_{\nu} \approx 0.42$ at higher frequencies \citep{gom12}. There 
could be a similar break for \pwn. If we fit a power-law to the radio data 
alone, the spectral index becomes $\alpha_{\nu,{\rm radio}} = 0.14\pm0.07$.
Similarly, we can also fit a power-law to the ACA data and the AKARI L15 data at 
$(1.92\pm0.36)\times10^{13}$~Hz, and the power-law for this range then is 
$\alpha_{\nu,{\rm mm/IR}} = 0.35^{+0.07}_{-0.05}$. As shown in
Figure~\ref{fig:multi2}, the two power-laws intersect at $\sim 8.4\EE{10}$~Hz,
i.e., somewhat higher frequency than the $\sim 3\EE{10}$~Hz break for the Crab. With
dual power-laws, the total steepening in power-law index becomes
$\Delta = 0.73^{+0.11}_{-0.12}$, which is again greater than $\Delta = 0.5$
at $\sim 2\sigma$ level.

However, the difference in power-law index between $\alpha_{\nu,{\rm mm/IR}}$
and $\alpha_{\nu,{\rm UVIR}}$ is $0.52\pm0.11$, which is fully consistent with a cooling
break of 0.5. Returning to the model of \citet[][see also \cite{Lyu19}]{bucc11}, there could be two 
populations of relativistic electrons, one being responsible for radio emission, and the other giving rise 
to the mm/IR part. The latter could experience a break due to synchrotron losses, which
is reflected in a power-law break of $\sim 0.5$ at $\nu_{\rm break} \sim 2\EE{13}$~Hz. If we use 
$B = 8\times10^{-4}$~G, as indicated from X-rays, $\tau_{\rm synch} \sim 190$~years,
which is markedly less than the age of the remnant, although not as low as 
$\tau_{\rm synch} \sim 2.6$~years estimated by \citet{Petre07}, from a presumed 
cooling break in X-rays instead of at $\sim 2\EE{13}$~Hz. Inspired by the model 
of \citet{Lyu19}, which suggests a cooling break at $\sim 0.01$~eV for the Crab PWN,
$\tau_{\rm synch} \sim 190$~years seems more likely than $\sim 3$ years for \pwn. 

%In any case, $\tau_{\rm synch} \ll 1\,100$~years probably requires that injection of relativistic 
%electrons can occur relatively far out in the PWN.
%A power-law with intermediate power-law index between roughly
%$9\EE{10} - 2\times10^{13}$~Hz for \pwn\ could be due to a spread in magnetic 
%field strength at injection sites in the nebula, or a spread in time since particle
%injection. 
%As a matter of fact, there is some evidence from optical polarization
%maps of presumably relatively recent activities along the PWN major axis of 
%\pwn\ \citep{nlun11}. 

%In Figure~\ref{fig:multi2}, we have assumed that the upper
%frequency of the intermediate power-law lies within the AKARI L15 band. This is
%lower than the likely high-energy break for the Crab PWN, which is at
%$\sim (1-2)\times10^{14}$~Hz \citep{tem06}. The idea of a dual break agrees with 
%the model used by \citet{Williams08} \citep[based on arguments
%by][]{rey09} which displays a gradual change in power-law index for \pwn\ around 
%$\sim 2\times10^{13}$~Hz. 

A dual spectral break in the radio-IR range for the synchrotron
emission in \pwn\ as shown in Figure~\ref{fig:multi2} allows for the possibility of dust
emission to explain excess emission at 24 $\mu$m. Such a dust
component was discussed by \citet{Williams08}, and in
Figure~\ref{fig:multi2} we include emission (shown in red, and highlighted with a black arrow) 
from $2.5\times10^{-3} \Msun$ of silicate dust with a temperature of 
$T_{\rm dust} = 55$~K (red solid line). The dust emission was calculated assuming
optically thin forsterite dust. The flux density $F_{\lambda}$ at wavelength 
$\lambda$ can be written as 
%\subsection{Maths}
%\label{sec:maths} % used for referring to this section from elsewhere
%
%Simple mathematics can be inserted into the flow of the text e.g. $2\times3=6$
%or $v=220$\,km\,s$^{-1}$, but more complicated expressions should be entered
%as a numbered equation:
%
\begin{equation}
    F_{\lambda}=\frac{\kappa_{\lambda} M_{\rm dust} B_{\lambda}(T_{\rm dust})}{D^2}~,
\label{eq:dustflux}
\end{equation}
where $M_{\rm dust}$ is the dust mass, $D$ the distance to \pwn\ 
(i.e., 50 kpc),  $B_{\lambda}(T_{\rm dust})$ the Planck function 
for temperature $T_{\rm dust}$, and $\kappa_{\lambda} \propto \lambda^{-\beta}$
the mass absorption coefficient of the selected type of dust
\citep[e.g.,][]{hil83,mat17}. The values of $\kappa_{\lambda}$ and
$\beta$ are temperature dependent. We have assumed small-sized dust
particles, and interpolated $\kappa_{\lambda}$ and
$\beta$ in temperature using the results reported by \citet{men98}. 
The choice of silicate dust is motivated by the fact
that \citet{noz03} find that silicates may dominate in supernova 
ejecta of massive progenitors, especially in mixed ejecta. 
\citet{Williams08} show that the presumed dust component increases 
in strength towards longer wavelengths up to the end
of their spectra at $\sim 37~\mu$m. This is consistent with the model
in Figure~\ref{fig:multi2} where the curve in red (dust +
synchrotron) peaks at $\sim 50~\mu$m. Our dust mass estimate is
consistent with the $\sim 3\times10^{-3} \Msun$ of silicate dust 
with $T_{\rm dust} = 50\pm8$~K calculated by \citet{Williams08}.
In Figure~\ref{fig:multi2} we include (in grey) a model with those
parameters. As can be seen, this model does not fit the integrated AKARI 
L24 flux density well. This is not surprising since the model of \citet{Williams08} was tuned
to fit Spitzer data. The inclusion of the AKARI L24 data pushes the 
derived dust temperature to a slightly higher value than estimated by 
 \citet{Williams08}.

As can be seen in Figure~\ref{fig:multi2}, the AKARI and Spitzer flux densities are 
larger than the extrapolation along the best optical--IR synchrotron fit.
This could indicate a dust contribution all the way up in frequency 
to the VLT/NACO bands, i.e., up to $\sim 1.4\EE{14}$~Hz. This would
require dust temperatures ranging between $\sim 50-60$~K, or lower,
and up to several hundred Kelvins. Although warm dust is seen in SNe
\citep[cf.][and references therein]{mat17}, 
detected PWN-embedded dust usually has temperatures lower than 
100~K \citep{chaw19}. 
SNR 1E 0102.2-7219 in the Small Magellanic Cloud (SMC) may be an exception 
at first glance. \citet{sands09} argue that small amounts 
($\sim 2\EE{-5} \msun$) of newly formed forsterite dust with 
$T_{\rm dust} \approx 140$~K is associated with the supernova ejecta, 
and a pulsar was recently reported for this remnant. However, no PWN
\citep{Vogt18} has yet been found, so SNR 1E 0102.2-7219 is probably more 
similar to Cas A than \pwn\ when it comes to the central source and its 
immediate surroundings.

A more likely reason for the high AKARI and Spitzer flux densities 
$\lsim 20 \mu$m is line contribution. For example, as shown in 
Figure~\ref{fig:S11}, the AKARI S11 band observations of \pwn\
\citep{Williams08} include [\ion{S}{IV}]~10.5~$\mu$m
at $f_{[\ion{S}{IV}]} \approx 7.3\EE{-14}~{\rm erg~cm}^{-2}~{\rm s}^{-1}$, and
[\ion{Ne}{II}]~12.8~$\mu$m at 
$f_{[\ion{Ne}{II}]} \approx 5.0\EE{-14}~{\rm erg~cm}^{-2}~{\rm s}^{-1}$. 
In Figure~\ref{fig:S11}, the blue solid line is part of the 
$\alpha_{\nu,{\rm UVIR}} = 0.87$ power-law fit in
Figures~\ref{fig:multi} and \ref{fig:multi2}, and the fluxes of the spectral lines are 
the broad lines (FWHM $\sim 1000~\kms$) reported by
\citet{Williams08}. Taking into account the transmission
of the AKARI S11 filter (cf. Figure~\ref{fig:S11}), the measured flux
in the spectral lines is $\sim 20\%$ of the continuum flux. 
If we add this estimated line contribution to the power-law
fit value for the continuum in the AKARI S11 band, which is
$3.81$~mJy at at $2.84\EE{13}$~Hz, the total flux density becomes
$4.57$~mJy. This agrees excellently  with the observed integrated flux
density in the AKARI S11 band which is $4.62\pm0.27$~mJy (cf. Table~\ref{tab:fluxtable}).

\begin{figure}
\includegraphics[width=8.9cm]{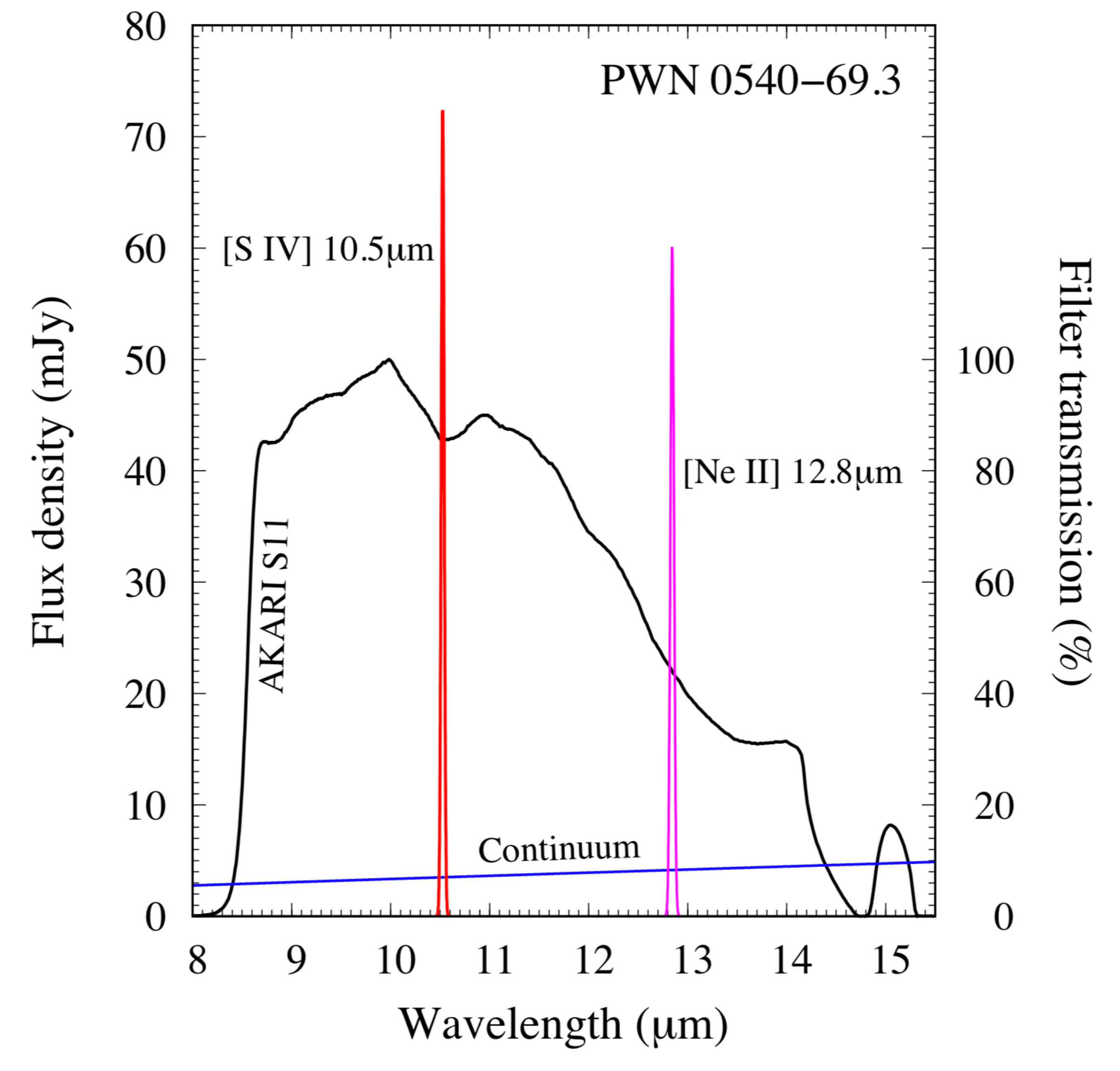}
\caption{Emission from \pwn\ in the AKARI S11 band. The blue solid line is a
power-law fit to the synchrotron continuum, adopting the 
$\alpha_{\nu,{\rm UVIR}} = 0.87$ power-law index from
Figures~\ref{fig:multi} and \ref{fig:multi2}. The two strongest spectral lines in this band, 
 [\ion{S}{IV}]~10.5~$\mu$m and [\ion{Ne}{II}]~12.8~$\mu$m, are also shown. 
Fluxes and widths of the lines are from \citet{Williams08}. The normalized filter 
transmission of the S11 band is highlighted by the black solid line. See text for
further details.}
\label{fig:S11}
\end{figure}

In our flux estimates of the continuum emission from \pwn, we have made
no correction for emission from \psr. \citet{serf04} made a detailed discussion
of this for the optical/UV part, and their results indicate that the pulsar would
contribute $\sim 5$\% if the pulsar were not spatially resolved. The
multi-wavelength compilation of \citet{serf04} also show that the PWN and the 
pulsar are of equal strength in X-rays (cf. Figure~\ref{fig:pwn_multii}), but that 
in radio the pulsar contribution is negligible compared to that from the PWN. This
trend is confirmed by \citet{Mignani12} whose results point to a $\sim 3$\% pulsar
contribution in the $H$ band. We can therefore safely ignore any pulsar contribution
to the integrated flux densities in the ACA, AKARI and Spitzer bands. As outlined before, uncertainties
due to spectral lines are larger. Further observations of \pwn\ in the near- and 
mid-infrared, preferably spectroscopic to remove the flux from spectral lines, 
should solve whether dust contributes to the emission measured by Spitzer and AKARI.
For example, the James Webb Space Telescope (JWST), with its expected spectral range
between $0.6-28.5~\mu$m and 0\farcs1 optical spatial resolution, is an ideal tool 
for this task.

In our analysis in Section~\ref{sec:fluxes}  we obtained 
$\alpha_{\nu,\text {SNR}} - \alpha_{\nu,\text {PWN}} = 0.47\pm0.03$ for the difference in spectral 
index between the remnant shell and the PWN for the spectral interval $8.6-145$ GHz. Using
Table~\ref{tab:fluxtable} we find $\alpha_{\nu,\text {PWN}} = 0.19{\pm0.04}$, and hence 
$\alpha_{\nu,\text {SNR}} = 0.64{\pm0.05}$. \citet{bra14} found $\alpha_{\nu,\text {SNR}} = 0.65{\pm0.01}$
for the range $1.4-8.6$ GHz, fully consistent with a constant spectral slope over two orders of magnitude
in frequency. Our value for the remnant shell is also consistent with typical values for synchrotron radiation
from other radio SNRs \citep{Green09}. It is also consistent with the spectrum for the synchrotron emission 
from the circumstellar ring around SN~1987A where $\alpha_{\nu,\text {87A}} = 0.70{\pm0.06}$ 
\citep{Cig19} for the same extended frequency interval we have discussed for \snr\ 
\citep[see also][]{Zan14}.

% Example table
%\begin{table}
%	\centering
%	\caption{This is an example table. Captions appear above each table.
%	Remember to define the quantities, symbols and units used.}
%	\label{tab:example_table}
%	\begin{tabular}{lccr} % four columns, alignment for each
%		\hline
%		A & B & C & D\\
%		\hline
%		1 & 2 & 3 & 4\\
%		2 & 4 & 6 & 8\\
%		3 & 5 & 7 & 9\\
%		\hline
%	\end{tabular}
%\end{table}

\section{Conclusions}
\label{sec:Conclusions}
We have observed \snr\ with the Atacama Compact Array (ACA) in Bands 4 (137--153 GHz) and 6 (213.5--233.5 GHz),
which is a new frequency range for this object. As the half-power beam-width (HPBW) of ACA is similar 
to the size of the remnants pulsar-wind nebula (PWN) in Band 6 ($7\farcs5 \times 5\farcs2$), and slightly larger than the 
PWN in Band 4 ($10\farcs8 \times 8\farcs3$), the PWN was only partially resolved. 

We also use deep radio observations 
obtained with the Australia Compact Array (ATCA) at 3 cm, with an HPBW of $\approx 0\farcs8$. We used the 3-cm data as 
a template for the emission in the millimeter-range, and smeared these data with the HPBWs of ACA Bands 4 and 6. 
These simulations suggest a similar overall structure at millimetre wavelengths compared to that at 3 cm. In particular, 
we recover the strong emission $\sim 1\farcs5$ to the south-west of the pulsar seen at other wavelengths.
If we include published radio flux densities  \citep{bra14}, we obtain a synchrotron spectrum  
$F_{\nu} \propto  \nu^{-0.17\pm{0.02}}$ for the frequency interval $1.449 - 233.50$~GHz. 

To draw conclusions about how the radio-millimetre wavelength range of the spectrum joins to the synchrotron spectrum at 
higher frequencies, and whether there could be dust in the PWN, we have evaluated published data in the UV, optical and IR,
as well as included previously unpublished AKARI IR data. We show that some of these data are seriously contaminated
by spectral line emission, and are therefore not suitable for analyzing the continuum emission. For the 
UV/IR part of the synchrotron spectrum, we find $F_{\nu} \propto \nu^{-0.87^{+0.10}_{-0.08}}$. The break between the 
radio--millimetre and UV/IR power-law occurs at $\nu_{\rm break} = (5.2^{+3.6}_{-2.7})\times10^{12}$~Hz, which can be
used to estimate the magnetic field strength, $B = (3.9^{+1.1}_{-0.7}) \times10^{-4}$~G, in the PWN, if the lifetime of
synchrotron-emitting electrons is as large as the remnant age.
This field strength is slightly higher than previous findings of \citet{man93b} and \citet{bra14}.

To explain the high observed flux from the PWN at $24~\mu$m, and at the same allow for a break in the synchrotron spectrum 
between radio and millimetre wavelengths as is seen in the Crab, we find that dust with a temperature of $\sim 50-60$~K is 
needed. For 55~K, the mass is $\sim 2.5\times10^{-3} \Msun$ for forsterite dust. The inferred break between radio and 
millimetre wavelenghts would  occur at $\sim 8\EE{10}$~Hz, with the spectral slope changing  from 
$-0.14\pm0.07$ in the radio, to $-0.35^{+0.05}_{-0.07}$ in 
the millimetre to far-IR range. The total change in power-law index based between radio and the optical/UV is $\Delta \sim 0.7$,
with $\Delta = 0.5$, as in the standard case of synchrotron losses in a homogeneous medium, being excluded at $\sim 2\sigma$
level, regardless of whether there is a spectral break at $\sim 8\EE{10}$~Hz, or not. 

However, there is a possible scenario for \pwn, which is inspired by a recent model by \citet{Lyu19} for the Crab PWN, and 
which can accommodate a cooling break of $\sim 0.5$. In this model our spectra reveal two populations of synchrotron-emitting 
relativistic electrons: one that emits in the radio with $F_{\nu} \propto \nu^{-0.14\pm0.07}$, and another that emits in the 
millimetre to UV range. In the millimetre to far-IR range, the latter component emits according 
to $F_{\nu} \propto \nu^{-0.35^{+0.05}_{-0.07}}$. Although shallower than $\propto \nu^{-0.5}$, uncertainties are large enough
for this component to be consistent with Fermi acceleration. 
The observed steepening in power-law index by $0.52\pm0.11$ to $F_{\nu} \propto \nu^{-0.87^{+0.10}_{-0.08}}$ for
$\nu \gsim 2\EE{13}$~Hz could be due to synchrotron cooling, as in the Crab. For a magnetic field strength of 
$B \approx 8\EE{-4}$~G, as indicated by recent X-ray studies \citep{Ge19}, the synchrotron cooling time is 
$\sim 190$~years.
%, which likely means that a global injection of synchrotron-emitting electrons in the PWN is needed.
%Such a large value for $\Delta$ could
%signal inhomogeneities in the PWN, or spread in time since relativistic particle injection.

For ACA Band 4, we clearly detect the supernova remnant shell of \snr, and we can constrain the spectrum of its
synchrotron emission. We find that its spectral slope is steeper than for the PWN. The spectral slope is $-0.64\pm{0.05}$ between
$8.6-145$~GHz, which agrees with previous estimates for radio alone, as well as standard values for other remnants, 
and the radio/millimetre spectrum of the ring of SN~1987A.

\section*{Acknowledgements}
We thank the anonymous referee for important comments. PL acknowledges support
from the Swedish Research Council. The work of YS was partially supported by the RFBR grant 16-29-13009.
This paper makes use of the following ALMA data: ADS/JAO.ALMA\#2017.1.01391.S. ALMA
is a partnership of ESO (representing its member states), NSF (USA) and NINS (Japan),
together with NRC (Canada), MOST and ASIAA (Taiwan), and KASI (Republic 
of Korea), in cooperation with the Republic of Chile. The Joint ALMA Observatory 
is operated by ESO, AUI/NRAO and NAOJ. The National Radio Astronomy Observatory
(NRAO) is a facility of the National Science Foundation operated under cooperative
agreement by Associated Universities, Inc. The Australia Telescope Compact Array
(ATCA) is part of the Australia Telescope National Facility which is funded by the 
Australian Government for operation as a National Facility managed by CSIRO. The 
ATCA data reported here were obtained under Program C014. This research has also
made use of the NASA Astrophysics Data System (ADS) Bibliographic Services.
Furthermore, this research has made use of the SVO Filter Profile Service
(http://svo2.cab.inta-csic.es/theory/fps/) supported from the Spanish MINECO 
through grant AYA2017-84089. M.M. acknowledges support from STFC Ernest Rutherford
fellowship (ST/L003597/1).

%%%%%%%%%%%%%%%%%%%%%%%%%%%%%%%%%%%%%%%%%%%%%%%%%%

%%%%%%%%%%%%%%%%%%%% REFERENCES %%%%%%%%%%%%%%%%%%

% The best way to enter references is to use BibTeX:

%\bibliographystyle{mnras}
%\bibliography{example} % if your bibtex file is called example.bib
% Alternatively you could enter them by hand, like this:
% This method is tedious and prone to error if you have lots of references
%\begin{thebibliography}{99}

%%%%%%%%%%%%%%%%%%%%%%%%%%%%%%%%%%%%%%%%%%%%%%%%%%

%%%%%%%%%%%%%%%%% APPENDICES %%%%%%%%%%%%%%%%%%%%%

%\appendix
%
%\section{Some extra material}
%
%If you want to present additional material which would interrupt the flow of the main paper,
%it can be placed in an Appendix which appears after the list of references.

%%%%%%%%%%%%%%%%%%%%%%%%%%%%%%%%%%%%%%%%%%%%%%%%%%

% Don't change these lines
\bsp	% typesetting comment
\label{lastpage}
\end{document}